\documentclass[12pt, draftclsnofoot, onecolumn]{IEEEtran}
\usepackage{graphicx,amssymb,amstext,amsmath}
\usepackage{amsthm}
\usepackage{color}
\usepackage{txfonts}
\usepackage{array}
\usepackage{subfigure}
\usepackage{algorithm}
\usepackage{algorithmicx}
\usepackage{algpseudocode}
\usepackage{multirow}
\usepackage[switch,pagewise]{lineno}
\theoremstyle{definition}

\theoremstyle{plain}

\newtheorem{theorem}{ \bf Theorem}
\DeclareMathOperator*{\argmax}{arg\,max}
\DeclareMathOperator*{\argmin}{arg\,min}
\newcommand{\bv}{\mathbf{v}}
\newcommand{\bu}{\mathbf{u}}
\newcommand{\bw}{\mathbf{w}}

\newcommand{\calU}{\mathcal{U}}
\newcommand{\calV}{\mathcal{V}}
\newcommand{\calN}{\mathcal{N}}

\newcommand{\calM}{\mathcal{M}}
\newcommand{\calS}{\mathcal{S}}
\newcommand{\rmR}{\mathrm{R}}
\newcommand{\rmD}{\mathrm{D}}
\newcommand{\bbR}{\mathbb{R}}
\newcommand{\bbN}{\mathbb{N}}

\usepackage[ colorlinks = true,
             linkcolor = blue,
             urlcolor  = blue,
             citecolor = red,
             anchorcolor = green,]{hyperref}

\usepackage{cite}
\usepackage{subfigure,transparent,color,graphicx} 

\allowdisplaybreaks[2]
\begin{document}
\title{Efficient User Scheduling for Uplink Hybrid Satellite-Terrestrial Communication}

\author{Lina Zhu, Lin Bai, Lin Zhou and Jinho Choi 
\thanks{L. Zhu is with the School of Communication Engineering, Hangzhou Dianzi University, Hangzhou, Zhejiang, China, 310018 (Email: zhulina@hdu.edu.cn). She was with the School of Electronic and Information Engineering, Beihang University, Beijing, China, 100083 (Email: zhulina@buaa.edu.cn)}
\thanks{L. Bai and L. Zhou are with the School of Cyber Science and Technology, Beihang University, Beijing, China, 100083 (Emails: \{l.bai, lzhou\}@buaa.edu.cn). They are also with the Beijing Laboratory for General Aviation Technology, Beihang University, Beijing.}
\thanks{J. Choi is with the School of Information Technology, Burwood, Deakin University, Australia (E-mail: jinho.choi@deakin.edu.au).}
}
\maketitle

\begin{abstract}
Due to increasing demands of seamless connection and massive information exchange across the world, the integrated satellite-terrestrial communication systems develop rapidly. To shed lights on the design of this system, we consider an uplink communication model consisting of a single satellite, a single terrestrial station and multiple ground users. The terrestrial station uses decode-and-forward (DF) to facilitate the communication between ground users and the satellite. The channel between the satellite and the terrestrial station is assumed to be a quasi-static shadowed Rician fading channel, while the channels between the terrestrial station and ground users are assumed to experience independent quasi-static Rayleigh fading. We consider two cases of channel state information (CSI) availability. When instantaneous CSI is available, we derive the instantaneous achievable sum rate of all ground users and formulate an optimization problem to maximize the sum rate. When only channel distribution information (CDI) is available, we derive a closed-form expression for the outage probability and formulate another optimization problem to minimize the outage probability. Both optimization problems correspond to scheduling algorithms for ground users. For both cases, we propose low-complexity user scheduling algorithms and demonstrate the efficiency of our scheduling algorithms via numerical simulations.
\end{abstract}


\IEEEpeerreviewmaketitle

\section{Introduction}\label{sec:intro}

To expand communication to a broader coverage area and meet the demand of higher throughput~\cite{survey_on_mobile,Robust_secure_beamforming}, satellite communication systems have been extensively studied and utilized across the world. Due to heavy shadowing caused by obstacles when the user moving into buildings or vegetation areas, the line-of-sight (LOS) link between the satellite and terrestrial users might be unavailable~\cite{Presence_of_Interference}. To combat the performance degradation, hybrid satellite-terrestrial relay systems are proposed to improve coverage and reliability of satellite communication with the aid of terrestrial relays~\cite{Energy_efficient,Outage_Analysis,Secrecy_Outage_Analysis}. Most current hybrid satellite-terrestrial systems adopt either amplify-and forward (AF) or decode-and-forward (DF)~\cite{Dual-Hop_Communication,Max-Max_User-Relay} relaying technique.

For a hybrid satellite-terrestrial system, the performance is strongly related to the design of multiuser access schemes~\cite{Max-Max_User-Relay},~relay selection methods~\cite{Relay_Selection_Outdated_CSI} and power allocation strategies~\cite{Improved_Performance}.
For satellite-terrestrial systems with a single relay, multiuser access schemes, especially user scheduling algorithms, are critical and should be optimized jointly with power allocation schemes. Furthermore, the development of integrated space-terrestrial information network (SIN) and the 6th Generation (6G) communication architecture enforces the system to allow access of massive number of devices. To meet the extremely high throughout and massive connectivity demands of future communication systems, it is vital to comprehensively study efficient user scheduling strategies with low complexity for hybrid satellite-terrestrial communication systems.

To shed lights on the design of such a system, we study user scheduling algorithms for the hybrid satellite-terrestrial model with a single satellite, a single terrestrial station and multiple ground users. The channel from the terrestrial station to the satellite is assumed to be a quasi-static shadowed Rician fading channel, and the channel from each user to the terrestrial station is assumed to be a quasi-static Rayleigh fading channel. Our choice of quasi-static fading is validated by information theoretical analysis of low-latency communication in~\cite{Multi-Connectivity,yang2014quasi}. In particular, under quasi-static fading channels, the asymptotic notion of outage capacity and outage probability provide tight approximations to the non-asymptotic performance at low latency. We choose shadowed Rician fading since it provides a good approximation for the land-mobile satellite communication channel~\cite{Land_Mobile_Satellite}. For the links between each ground user and the relay, we choose Rayleigh fading because it is widely used in wireless communications, and is accurate for most terrestrial environment with rich scatters~\cite{Fundamental}.

\subsection{Main Contributions}

We study the hybrid satellite-terrestrial model with either instantaneous channel state information (CSI) or channel distribution information (CDI) available at the terrestrial station  and the satellite. When instantaneous CSI is available, the terrestrial station is informed of the exact channel fading levels of each user, and the satellite is informed of the exact fading level from the terrestrial station. When CDI is available, the terrestrial and the satellite are only provided with the distribution of respective fading models. We study user scheduling algorithms for both cases. Our contributions are summarized as follows.

When instantaneous CSI is available, combining capacity results for the multiple access channel and the relay channel~\cite{IF_the}, we derive the optimal instantaneous achievable rate region of all ground users. Subsequently, under a homogeneous target rate constraint,  we formulate an optimization problem to maximize the achievable sum rate via user scheduling algorithms. The optimal solution requires an exhaustive search and is too complicated for practical use. As a compromise, we first derive theoretical upper and lower bounds on the maximum achievable rate and then propose low-complexity scheduling algorithms. In particular, we propose a greedy algorithm that achieves close-to-optimal performance and a further simplified sub-optimal algorithm that has almost linear complexity. The efficiency, computational complexity and stability of our scheduling algorithms are analytically demonstrated and numerically verified.

When CDI is available, the instantaneous achievable rate region can not be obtained. Instead, under a homogeneous target rate constraint for all ground users, we derive the outage probability of the uplink communication as a function of the rate constraint and CDI. Subsequently, we formulate an optimization problem to minimize the outage probability and derive a theoretical lower bound that serves as the benchmark. Furthermore, to obtain a feasible solution to the optimization problem, we propose a user scheduling algorithm using the idea of alternative optimization (AO)~\cite{AO1,AO2}. Via numerical simulations, we find that the outage probability obtained from our user scheduling algorithm converges to our derived benchmark after a small number of iterations. We also numerically illustrate the optimality and computational complexity of our scheduling algorithm by comparing with the optimal exhaustive search method.

\subsection{Related works}

The study on hybrid satellite-terrestrial communication systems has a long history. In contrast to our setting, most papers focused on the downlink model, where a satellite transmits messages to a group of ground users, e.g.,~\cite{Ergodic_Capacity_of,On_the_Performance,CCI_andOutdated_CSI,Joint_Beamforming_and_Power_Allocation,Non-Orthogonal_Multiple_Access_Based_Integrated}. In~\cite{Ergodic_Capacity_of}, for a multiuser hybrid AF satellite-terrestrial cooperative network, Yuan \emph{et al.} proposed an opportunistic user scheduling algorithm using signal-to-noise ratio (SNR)-threshold based feedback. In~\cite{On_the_Performance}, for a multiuser AF hybrid satellite-terrestrial relay network with opportunistic scheduling, Kang \emph{et al.} derived the analytical bounds on the ergodic capacity. For a multiuser satellite-terrestrial downlink model with a multi-antenna satellite and a single-antenna AF relay, Bankey and Upadhyay~\cite{CCI_andOutdated_CSI} analyzed the ergodic capacity for opportunistic user scheduling with outdated CSI. However, the spectrum efficiency of the orthogonal opportunistic user scheduling algorithms of the above studies is very low since only one user is allowed to access the channel at a time slot. To improve the throughput and spectrum efficiency of the downlink hybrid system, subsequent works apply non-orthogonal multiple access (NOMA) into hybrid satellite-terrestrial systems to support multiple users simultaneously in a time slot, e.g.,~\cite{Joint_Beamforming_and_Power_Allocation,Non-Orthogonal_Multiple_Access_Based_Integrated}. However, studies on downlink NOMA-based integrated terrestrial-satellite network are significantly different from our uplink analysis in this paper.

The more related works to ours is the studies of the uplink hybrid satellite-terrestrial system, e.g.,~\cite{Closed-Form_Capacity,Massive_Access,Two-Way_AF_Satellite}. In~\cite{Closed-Form_Capacity}, Yang \emph{et al.} derived an upper bound on the ergodic capacity for a multi-beam geostationary earth orbit mobile satellite communication system. In~\cite{Massive_Access}, Huang \emph{et al.} proposed a space division multiple access (SDMA) scheme to maximize the average sum rate of a hybrid satellite-terrestrial AF relaying network, where a high-altitude platform is deployed as a relay to assist the transmission from ground users to the satellite. In~\cite{Two-Way_AF_Satellite}, Arti studied a two-way AF satellite system with only two ground users and one terrestrial relay and designed beamforming and combining vectors for the satellite.

However, none of the above works use Shannon theory~\cite{shannon1948mathematical} to study the critical problem of efficient user scheduling algorithms to achieve the full potential of the uplink hybrid satellite-terrestrial system. Although above works considered multiple antennas, the analysis is mostly achievable without optimality guarantee. Specifically, these studies used dimensional advantage of multiple antennas to improve the throughout with SDMA or a certain beamforming design. Note that the capacity is the maximal rate achievable of any communication scheme and analysis based on specific coding scheme without optimality guarantee is far from satisfactory. In this paper, to complement the existing literature in the uplink hybrid satellite-terrestrial system, we use information theoretical results to study efficient user scheduling algorithms and demonstrate close-to-optimality performance of our proposed algorithms. For simplicity, we present results for the single antenna terminals in this paper. Our results can be generalized to the multi antenna terminals by using corresponding information theoretical results for the multiple input and multiple out channels~\cite{Wang2005relay,goldsmith2003jsac}.

\section{Problem Formulation}
\subsection*{Notation}
We use $\bbR$, $\bbR_+$ and $\bbN$ to denote the set of real numbers, positive numbers and integers respectively.
Denote by $\mathcal{A}\setminus\mathcal{B}$ the minus between set $\mathcal{A}$ and $\mathcal{B}$.
For any $x\in\bbR$, we use $\lfloor x\rfloor$ to denote the maximum integer that is less  than or equal to $x$. For any $(a,b)\in\bbN^2$, we use $[a:b]$ to denote the set of integers between $a$ and $b$ and use $[a]$ to denote $[1:a]$. Furthermore, for any two integers $(a,b)\in\mathbb{N}^2$ such that $a<b$, denote by $[b]_a$ the set of vectors that contain exactly $a$ unique elements of $[b]$ and by $\mathcal{F}(a,b)$ the set of bijective mappings from $[b]_a$ to $[b]_a$.

\subsection{The Uplink Communication Model}

\begin{figure}
\centering
\includegraphics[scale=0.4]{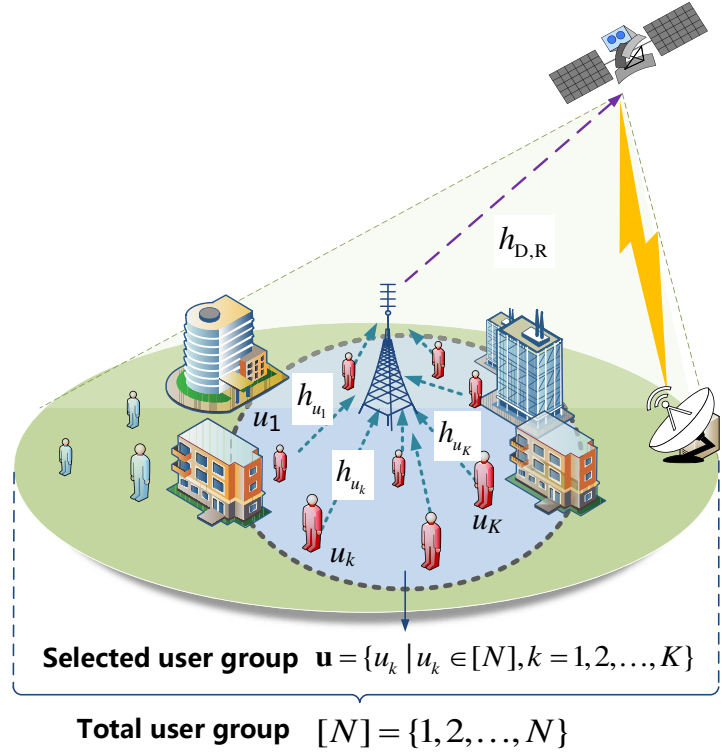}\\
\caption{System model of the hybrid satellite-terrestrial communication network.}
\label{system_model}
\end{figure}

We consider an uplink model for the hybrid satellite-terrestrial communication system as shown in Fig. \ref{system_model}. There are $N$ ground users with indices $\mathcal{N}=[N]$, one terrestrial relay, and one satellite, all assumed to have a single antenna. Each user wishes to transmit messages to the satellite with the assist of the relay. Since the resource of the satellite is finite, we assume that at most $K$ users are allowed to transmit messages in a time slot simultaneously. Let $\bu=\{u_1,\ldots,u_K\}\in[N]_K$ denote the vector of the indices of the selected users by a scheduling algorithm.

At each time slot, after user scheduling, the communication between the selected $K$ users and the satellite proceeds in the following two phases. In the first phase, for each $k\in[K]$, the selected user $u_k$ aims to send a message $M_{u_k}\in[1:2^{nR_{u_k}}]$ to the relay with power $P_1$, rate $R_{u_k}$ and blocklength $n$. Each message $M_{u_k}$ is then encoded into a length-$n$ codeword $X_{u_k}^n$. The received signal at the terrestrial relay is
\begin{equation}\label{rec_relay}
  Y_\mathrm{R}^n=\sum_{k\in[K]}h_{u_k}X_{u_k}^n+Z_{u_k}^n,
\end{equation}
where $h_{u_k}$ is the channel fading coefficient between the user $u_k$ and the relay, and $Z_{u_k}^n$ is the independent additive white Gaussian noise (AWGN) with zero mean and unit variance. Using $Y_\mathrm{R}^n$ and CSI for all $K$ users, the terrestrial relay decodes the messages of each user as $\{\tilde{M}_{u_k}\}_{k\in[K]}$. In the second phase, the relay encodes its decoded messages $\{\tilde{M}_{u_k}\}_{k\in[K]}$ into a combined codeword $X_\mathrm{R}^n$ with power $P_2$ and sends it to the satellite. The received signal at the satellite is
\begin{align}
\label{rec_D}
  Y_\mathrm{D}^n=h_{\mathrm{D},\mathrm{R}}X_\mathrm{R}^n+Z_{\mathrm{D},\mathrm{R}}^n,
\end{align}
where $h_{\rmD,\rmR}$ is the channel fading coefficient between the relay and the satellite, and $Z_{\rmD,\rmR}^n$ is the independent AWGN with zero mean and unit variance. Finally, using the noisy outputs $Y_\rmD^n$ and CSI on the fading coefficient $h_{\rmD,\rmR}$, the satellite decodes the message $\{\hat{M}_{u_k}\}_{k\in[K]}$.

\subsection{Fading Channels and Assumptions}
We assume that the channel between each ground user $i\in[N]$ and the terrestrial relay is an AWGN with Rayleigh fading, i.e., the fading coefficient satisfies $h_{i}\sim \mathcal{CN}(0,2\sigma_{i}^2\mathbf{I})$. The fading random variables for different ground users are independent. Furthermore,  we assume that the channel gain $|h_{\rmD,\rmR}|$ for the channel from the relay to the satellite follows a Shadowed-Rician (SR) distribution with parameters $(\Omega,b_0,m_s)$, i.e., $|h_{\rmD,\rmR}| \sim \mathrm{SR}(\Omega, b_0, m_s)$, where
 $\Omega$, $b_0$ and $m_s$ are positive real numbers.
Here, $\Omega$ denotes the the average power of the LOS component, $m_s$ is the fading order, and $2b_0$ represents for the average power of the scattered component.

To evaluate the fundamental limits of the uplink hybrid satellite-terrestrial system, depending on the availability of CSI, we study  the instantaneous sum rate as well as the outage probability and formulate corresponding optimization problems to schedule the transmission of ground users. Specifically, we consider the following two cases.
\begin{itemize}
\item {\bf{Case 1: Perfect Instantaneous CSI}}

Similar to most existing works, e.g., \cite{Joint_beamforming,terrestrial_5G_networks}, we assume that perfect CSI of all channels are available. The CSI can be obtained by channel estimation during each transmission time slot via a backhaul channel~\cite{Joint_Beamforming_and_Power_Allocation}. In this setting, we derive the achievable instantaneous sum rate and propose efficient user scheduling algorithms to maximize the sum rate.

\item {\bf{Case 2: Statistical CSI}}

The assumption of perfect instantaneous CSI can be impractical for practical satellite communication systems. To address this concern, we also study the case where only CDI is available. Specifically, the relay only knows the distribution of fading coefficients for channels from each ground user and the satellite only knows the distribution of the fading coefficient for the channel from the relay. In practical systems, CDI can be estimated via channel parameter estimation methods, e.g., the MUSIC algorothm in~\cite{MUSIC}. Under this setting, we derive the outage probability and propose user scheduling algorithms to minimize the outage probability.
\end{itemize}

\section{Main Results for the Case with Perfect Instantaneous CSI}\label{performace_case1}
In this section, we analyze the instantaneous achievable rate for the given system and formulate an optimization problem to maximize the sum rate. A solution to the optimization problem is equivalent to a user scheduling algorithm. The optimal solution requires an exhaustive search, which is highly complicated and renders its impossible for practical communication systems. To solve this problem, we first derive theoretical lower and upper bounds and then propose low-complexity user scheduling algorithms that can exploit the tradeoff between the complexity and performance.

\subsection{Analysis of the Instantaneous Achievable Sum Rate}
\label{Problem Formulation sum rate}

For each $k\in[K]$, let $S_{u_k}:=P_1|h_{u_k}|^2$ denote the received SNR for the channel between user $u_k\in[N]$ and the relay and let $S_{\mathrm{D},\mathrm{R}}:=P_2|h_{\mathrm{D},\mathrm{R}}|^2$ denote the SNR for the channel between the relay and the satellite. Using cut-set bounds for the relay channel and the multiple access channel in~\cite[Sec. 19.1 and Theorem 19.1]{IF_the}, we conclude that given selected users $\mathbf{u}=(u_1,\ldots,u_K)$, the instantaneous rate of each user $R_{u_j}$ is upper bounded by
\begin{align}\label{sumrate}
  \sum_{j\in[K]}R_{u_j}\leq \min\left\{ C\left(\sum_{j\in[K]}S_{u_j}\right), C(S_{\mathrm{D},\mathrm{R}})\right\},
\end{align}
where $C(a)=\log(1+a)$ is the capacity of an AWGN channel with SNR $a\in\mathbb{R}_+$.

In a full-duplex DF relaying system with perfect CSI, the sum rate upper bound given by \eqref{sumrate} can be achieved via block Markov coding and backward decoding (\cite[Sec. 16.4.4 and Table 16.2]{IF_the}), when superposition encoding and successive cancellation decoding are used at the relay and the satellite. For subsequent analysis, we consider the case where the rate tuples $\{R_{u_k}\}_{k\in[K]}$ lies at a corner point on the rate region and thus maximal sum rate is achieved.
In particular, let $\mathbf{u}:=(u_1,\ldots,u_K)\in[N]_K$ and let $u_1\rightarrow u_2\rightarrow \ldots\rightarrow u_K$ be the decoding order of successive cancellation decoding at the relay, then the signal-to-interference plus noise ratio (SINR) for user $u_k$ at the relay satisfies
\begin{equation}\label{rec_relay_SNR}
  \gamma_{u_k}=\frac{S_{u_k}}{\sum_{i\in[k+1:K]}S_{u_i}+1}, k\in[K-1]\mathrm{~and~}\gamma_{u_K}=S_{u_K}.
\end{equation}
Since the decoding order of successive cancellation at the satellite can be different from that at the relay, we use a bijective mapping $\Psi\in\mathcal{F}(K,N)$ and the decoding order $\mathbf{u}$ at the relay to describe the decoding order at the satellite. Specifically, for any $k\in[K]$, when $u_k$ is the $k$-th decoded message at the relay, then it is the $\Psi(u_k)$-th decoded message at the satellite. Thus, the decoding order $\mathbf{u}$ and the bijective mapping $\Psi\in\mathcal{F}(K,N)$ suffice to describe the set of selected users and the decoding orders at both the terrestrial station and the satellite. Note that the transmission power $P_2$ is split among all $K$ messages and thus the SINR for each message from the relay to the satellite satisfies
\begin{equation}\label{rec_relay_SNR2}
\begin{split}
  \gamma_{\Psi(u_k)}&=\frac{\alpha_{\Psi(u_k)}}{\sum_{i\in[\Psi(u_k)+1:K]}\alpha_i+1/S_{\mathrm{D},\mathrm{R}}},~\forall~k\in[K]\mathrm{~s.t.~}\Psi(u_k)\neq K \\ \gamma_{\Psi(u_k)}&=\alpha_{\Psi(u_k)}S_{\mathrm{D},\mathrm{R}}, \Psi(u_k)= K,
  \end{split}
\end{equation}
where $\alpha_{\Psi(u_k)}$ is the power splitting ratio of the message from user $u_k$ and satisfies $\sum_{k\in[K]}\alpha_{\Psi(u_k)}=1$.
Therefore, given $\mathbf{u}$ and $\Psi$, the instantaneous achievable rate of user $u_k$ satisfies
\begin{equation}\label{Individual}
  R_{u_k}=\min\left\{C(\gamma_{u_k}),C(\gamma_{{\Psi(u_k)}})\right\},~k\in[K].
\end{equation}

\subsection{Maximize Instantaneous Achievable Sum Rate via An Optimization Problem}
Under the assumption that at most $K$ users are allowed to transmit in a time slot, given a per user rate constraint $R_k\in\mathbb{R}_+$, we formulate the optimization problem in \textbf{Problem 1} to maximize the achievable instantaneous sum rate as follows:
\begin{align}
\mathrm{\bf Problem~1:}\quad R_{\mathrm{sum}}^* :=&\max_{\substack{\mathbf{u}\in([N]_K),\Psi\in\mathcal{F}(K,N),\\
 (\alpha_{\Psi(u_1)},\ldots,\alpha_{\Psi(u_K)})\in[0,1]^K}}\min\left\{ C\left(\sum_{j\in[K]}S_{u_j}\right), C(S_{\mathrm{D},\mathrm{R}})\right\}\label{objective}\\
&\mathrm{s.t.}~
R_{u_k}\geq R_{\mathrm{target}}\label{cp1},\\
&\quad\sum_{k\in[K]}\alpha_{\Psi(u_k)}=1\label{cp2},\\
&\quad S_{u_i}\geq S_{u_j}~\mathrm{for~}(i,j)\in[K]_2\mathrm{~and~}i>j\label{cp3}.
\end{align}
The objective function of \textbf{Problem 1} is the maximum achievable sum rate and we need to optimize over all possible schedule user decoding orders $\mathbf{u}$ at the terrestrial station, all possible decoding orders at the satellite determined by $\mathbf{u}$ and the bijective mapping $\Psi$, all power allocation vectors $(\alpha_{\Psi(u_1)},\ldots,\alpha_{\Psi(u_K)})$ at the relay. The first constraint in \eqref{cp1} makes sure that every scheduled user satisfies the rate constraint $R_{\mathrm{target}}$, the second constraint in \eqref{cp2} is the power constraint at the relay to make sure the total transmission power at the relay is $P_2$, and the third constraint in \eqref{cp3} guarantees that the decoding order at the relay: $u_1\rightarrow u_2\rightarrow\ldots\rightarrow u_K$ is in the descend order of received SINRs. To solve the problem, we need to jointly optimize the scheduled users $\mathbf{u}$ at the terrestrial station, the decoding order $\{\Psi(u_k)\}_{k\in[K]}$ at the satellite and power allocation vector $\{\alpha_{\Psi(u_k)}\}_{k\in[K]}$ at the relay.

We remark that \textbf{Problem 1} can be decomposed into two sub-problems. This is because $C(\sum_{u_j\in \mathcal{U}}S_{u_j})$ and $C(S_{\rmD,\rmR})$ depend on different optimized parameters. Specifically, we have sequential sub-problems \textbf{Problem 2(a)} and \textbf{Problem 2(b)}, where \textbf{Problem 2(b)} uses the solution of \textbf{Problem 2(a)} to proceed.
\begin{figure*}[ht]
\begin{minipage}{.53\linewidth}
\begin{align}
\nonumber&\mathrm{\bf Problem 2(a):}\\
&\max_{
     \mathbf{u}\in[N]_K} C\left(\sum_{u_k\in \mathcal{U}}S_{u_k}\right)\\
     &\mathrm{s.t.}~C(\gamma_{u_k})\geq R_{\rm{target}},~k\in[K],\label{ratecons1_1}\\
     &~\quad S_{u_i}\geq S_{u_j}~\mathrm{for~}(i,j)\in[N]_2\mathrm{~and~}i>j.\label{ratecons1_2}
\end{align}
\end{minipage}
\begin{minipage}{.45\linewidth}
\begin{align}
\nonumber&\mathrm{\bf Problem 2(b):}\\
&\max_{\substack{\Psi\in\mathcal{F}(K,N),(\alpha_{\Psi(u_1)},\ldots,\alpha(\Psi_(u_K)))\in[0,1]^K}} C(S_{\rmD,\rmR})\\
&\mathrm{s.t.~}C(\gamma_{\Psi(u_k)})\geq R_{\mathrm{target}}, k\in[K],\label{ratecons2}\\
&\quad~\sum_{k\in[K]}\alpha_{\Psi(u_k)}=1.
\end{align}
\end{minipage}
\hrulefill
\end{figure*}

By solving \textbf{Problem 2(a)}, we obtain the set of scheduled users and its decoding order $\mathbf{u}$ at the relay that maximizes the sum rates of $K$ users satisfying the rate constraint $R_{\mathrm{target}}$. Given $\mathbf{u}$, we can then solve \textbf{Problem 2(b)} for power allocation at the relay to maximize the achievable rate from the relay to the satellite that satisfies $R_{\mathrm{target}}$ for each scheduled user $u_k$.
Note that the objective function of $\textbf{Problem 2(b)}$ only depends on the fading coefficient $h_{\mathrm{\rmD,\rmR}}$ between the satellite and the relay and is thus independent from user scheduling and power allocation strategies.

With the constraint in \eqref{ratecons2}, the power allocation vector $\{\alpha_{\Psi(u_k)}\}_{k\in[K]}$ satisfies $\frac{\alpha_{\Psi(u_k)}}{\sum_{i=\Psi(u_k)+1}^{K}\alpha_i+1/S_{\mathrm{D},\mathrm{R}}}\geq 2^{R_\mathrm{target}}-1$
for $\Psi(u_k)\in[K-1]$ and $S_{\mathrm{D},\mathrm{R}}\alpha_K\geq 2^{R_\mathrm{target}}-1$. Thus, $\alpha_{\Psi(u_k)}\geq \frac{2^{R_\mathrm{target}}-1}{S_{\mathrm{D},\mathrm{R}}}2^{R_\mathrm{target}(K-\Psi(u_k))}$.
Combined with the power constraint that $\sum_{k\in[K]}\alpha_{\Psi(u_k)}=1$, we conclude that \textbf{Problem 2(b)} is feasible and we can always find a feasible power allocation vector $\{\alpha_{\Psi(u_k)}\}_{k\in[K]}$ if the number of users supported with rate $R_\mathrm{target}$ satisfies
\begin{equation}\label{Maximum_K_1}
  K\leq \frac{\log(S_{\mathrm{D},\mathrm{R}}+1)}{R_\mathrm{target}}.
\end{equation}
Note that (\ref{Maximum_K_1}) also implies that $S_{\mathrm{D},\mathrm{R}}$ should be larger than $2^{R_\mathrm{target}}-1$ to make sure that at least one user can be served with desired rate, i.e., $K\geq 1$.

\subsection{Upper and Lower Bounds on Instantaneous Achievable Sum Rate $R_{\mathrm{sum}}^*$}
\label{Upper and Lower Bounds}
To obtain an exact solution of \textbf{Problem 1}, we need to use an exhaustive search method over all possible user scheduling $\mathcal{U}\in[N]_K$ and find the largest achievable sum rate $R_{\mathrm{sum}}^*$. However, such a method is infeasible, especially when the total number of users $N$ is large. As a compromise, we derive closed-form upper and lower bounds on the instantaneous achievable sum rate $R_{\mathrm{sum}}^*$, which can be used as benchmarks for low complexity user scheduling algorithms.

We assume that \eqref{Maximum_K_1} holds so that \textbf{Problem 2(b)} is feasible and the achievable rate between the relay and the satellite is $C(S_{\rmD,\rmR})$. Then, to bound $R_{\mathrm{sum}}^*$, it suffices to derive upper and lower bounds for the optimal value of \textbf{Problem 2(a)}. For any decoding order $\bu\in[N]_K$, let $S_{\mathrm{max}}:=\max_{k\in[N]}S_{k}$ and $S_{\mathrm{min}}:=\min_{k\in[N]}S_{k}$. Furthermore, let $\calS=\{S_i\}_{i\in[N]}$, given any $a\in\bbR_+$ and any set $\mathcal{A}$, let $\phi(a,\mathcal{A}):=\min_{S_{i}\in \mathcal{A}:S_{i}\geq a} S_{i}$. Finally, given any $R_{\mathrm{target}}\in\mathbb{R}_+$, define the following quantities
\begin{align}
\gamma_{\mathrm{target}}&:=2^{R_{\mathrm{target}}}-1,\label{def:gamma}\\
  \hat{S}_{u_K}^{\mathrm{lb}}&:=\phi\left(\gamma_{\mathrm{target}},\mathcal{S}\right),
  ~\hat{S}_{u_k}^{\mathrm{lb}}:=\phi\left(\gamma_{\mathrm{target}}\left(\sum_{i\in[k+1:K]}\hat{S}_{u_i}^{\mathrm{lb}}+1\right)
  ,\mathcal{S}\setminus\{\hat{S}_{u_j}^\mathrm{lb}\}_{j\in[k+1,K]}\right),k\in[K-1]\label{fsp1a},\\
 S_{u_k}^\mathrm{ub}
 &:=\frac{S_{\mathrm{max}}}{(\gamma_{\mathrm{target}}+1)^{k-1}},k\in[K-1],~S_{u_K}^\mathrm{ub}:=\frac{S_{\mathrm{max}}}{(\gamma_{\mathrm{target}}+1)^{K-2}\gamma_{\mathrm{target}}}-1\label{UPB}
\end{align}

\begin{theorem}
\label{bounds}
Given any target rate $R_{\mathrm{target}}$ and corresponding $\gamma_{\mathrm{target}}$, the following results hold.
\begin{enumerate}
\item \textbf{Problem 2(a)} is feasible if and only if
\begin{align}
\label{claim1}
 \gamma_{\mathrm{target}}/S_{\mathrm{max}}\leq\min\left\{1,1/\left(\sum_{i\in[k+1:K]}\hat{S}_{u_i}^{\mathrm{lb}}+1\right)\right\}
\end{align}
\item If \textbf{Problem 2(a)} is feasible, then
\begin{align}
R_{\mathrm{sum}}^*\geq \min \left\{C\left(\sum_{k=1}^KS_{u_k}^\mathrm{lb}\right),C(S_{\mathrm{D},\mathrm{R}})\right\},
\end{align}
where $S_{u_k}^{\mathrm{lb}}=\hat{S}_{u_k}^{\mathrm{lb}}$ for $k\in [2,K]$ and $S_{u_1}^{\mathrm{lb}}=S_{\mathrm{max}}$;
\item $R_{\mathrm{sum}}^*$ is upper bounded by
\begin{align}
R_{\mathrm{sum}}^*\leq \min\left\{ C\left(\sum_{k=1}^KS_{u_k}^\mathrm{ub}\right), C(S_{\mathrm{D},\mathrm{R}})\right\}.
\end{align}
\end{enumerate}
\end{theorem}
The proof of \textbf{Theorem \ref{bounds}} is available in Appendix \ref{pf:bounds}.

\subsection{A Greedy Iterative User Scheduling Algorithm}
\label{User Scheduling_ins}

To efficiently solve \textbf{Problem 2(a)} with low complexity, we propose a user scheduling algorithm that corresponds to a solution of \textbf{Problem 2(a)} and achieves a near optimal performance of exhaustive search. Specifically, our algorithm proceeds in two phases: given CSI, we first derive the maximum number of users $K$ that can be served by the system and then propose a greedy iterative user scheduling algorithm to select $K$ users to transmit. Details are as follows.

\subsubsection{Determine the maximum number of supported users $K$}
Given an $R_{\mathrm{target}}$ and the terrestrial SNR values $\mathcal{S}$, we should first determine the number of supported users. This is because to ensure reliable communication of a multiple-access system, individual rate decreases with the increasing of user numbers \cite{IF_the}. In previous analysis, we derive an upper bound on $K$ in (\ref{Maximum_K_1}) that depends on the SNR $S_{\mathrm{D,R}}$ of the satellite-relay link. However, $K$ is also related with channel qualities of users. Combined with effects of both, we determine $K$ via the following steps:
\begin{enumerate}
  \item[Step 1:] Initialize $K=\min\left\{N,\left\lfloor\frac{\log(S_{D,R}+1)}{R_\mathrm{target}}\right\rfloor\right\}$.
  \item[Step 2:] Given $K$, if \eqref{claim1} in \textbf{Theorem \ref{bounds}} holds, end the algorithm; otherwise, reduce the value of $K$ by $1$ and repeat Step 2 until $K=0$.
\end{enumerate}

\subsubsection{Schedule users via a greedy iterative scheduling algorithm}

Note that $K$ has be chosen to ensure that \textbf{Problem 2(a)} is feasible.
Solving \textbf{Problem 2(a)} is then equivalent to proposing a user scheduling algorithm based on SNR values of all terrestrial channels. However, \textbf{Problem 2(a)} is a combinatorial optimization problem (COP), which is NP hard and may not have an analytical expression for optimal solution. Furthermore, it is not trivial to solve the problem using traditional integer programming solving methods such as dynamic programming or branch and bound \cite{IP}, because it is hard to determine the feasible region. To proceed, we propose a greedy algorithm that iteratively selects users and achieves a near optimal performance.

We need several definitions to describe our scheduling algorithm.
Given $\{S_{i}\}_{i\in[N]}$ of all terrestrial channels, let $\{S_{i}^{\uparrow}\}_{i\in[N]}$ be the increasing order and for each $n\in[N]$, let $L_{\mathrm{min}}(n)$ is the sum of the weakest $n$ SINRs of users, i.e., $L_{\mathrm{min}}(n):=\sum_{i\in[n]}S_{i}^{\uparrow}$.
Furthermore, given target rate $R_{\mathrm{target}}$ and the corresponding $\gamma_{\mathrm{target}}=2^{R_{\mathrm{target}}}-1$, define a sequence $\mathbf{T}=(T_1,\ldots,T_K)$ where
\begin{align}
T_1=\frac{S_{\mathrm{max}}}{\gamma_{\mathrm{target}}},~T_2=\min\left\{T_1-S_{u_2^*},\frac{S_{u_2^*}}{\gamma_{\mathrm{target}}}\right\},
~T_{k-1}=\min\left\{T_{k-2}-S_{u_{k-1}^*},\frac{S_{u_{k-1}^*}}{\gamma_{\mathrm{target}}}\right\},~k\in[3:K].
\end{align}

Our greedy user selection algorithm then operates as follows. For each $k\in[K]$, let $\calU_k$ denote the set of users that can be selected. To select the first user, we have $\calU_1=[N]$ and we select the first user $u_1^*$ such that its received SNR equals $S_{\mathrm{max}}$, i.e., $u_1^*=\argmax_{i\in[N]}S_{i}$. For the subsequent user selection, given each $k\in[K]$, let
\begin{align}
\calV_k&:=\big\{i\in\calU_k:~S_{i}\in[S_{\mathrm{min}},\min\{T_{k-1}-L_{\mathrm{min}}(k)-1,S_{u_{k-1}^*}\}]\big\}\label{def:vK}.
\end{align}
The $k$-th user $u_k^*$ is then selected as $u_k^*=\argmax_{i\in\calV_k}S_{i}$,
and $\calU_k=\calU_{k-1}\setminus\{u_k^*\}$. To ensure that we can select all $K$ users, we need to consider whether $\calV_k$ is empty for any $k\in[K]$. If $\calV_k$ is empty, we update $\calU_{k-1}$ as $\calU_k$ and trace back to $k-1$ to select another user other than $u_{k-1}^*$. Note that this is possible since after choosing $u_{k-1}^*$, we have already updated $\calU_{k-1}$.
The user selection stops until we select all $K$ users $\bu^*=(u_1^*,u_2^*,\ldots,u_K^*)$.

We summarize our greedy iterative user scheduling (GIUS) scheme in \textbf{Algorithm \ref{Iteration}}. Our simulation results demonstrate that in most cases, GIUS has performance close to that of exhaustive search while the computational complexity is significantly reduced.

\begin{algorithm}[ht]
\caption{Greedy iterative user scheduling algorithm (GIUS)}
\label{Iteration}
\begin{algorithmic}
\Require
$N$, $K$, terrestrial channel coefficients $\{h_{k}\}_{k\in[K]}$, transmission power $P_1$, target rate $R_{\mathrm{target}}$
\Ensure Scheduled users $\bu^*=\{u_1^*,u_2^*,\ldots,u_K^*\}$.
\State $\calU_1=[N]$; Calculate $S_{i}$ for $i\in[K]$, and $S_{\mathrm{max}}$, $S_{\mathrm{min}}$, $\gamma_{\mathrm{target}}$.
\State Choose the first user as $u_1^*=\argmax_{i\in[N]}S_{i}$, $k\gets 2$,
\State $\calU_2=\calU_1\setminus\{u_1^*\}$,

\While{$k\leq K$}
\State Calculate the set $\calV_k$ from \eqref{def:vK}
\If {The set $\calV_k$ is empty}
\State $k\gets k-1$
\Else
\State Select the $k$-th user as $u_k^*=\argmax_{i\in\calV_k}S_{i}$
\State $\calU_k\gets \calU_k\setminus\{u_k^*\}$, $\calU_{k+1}\gets \calU_k$
\State $k\gets k+1$
\EndIf
\EndWhile
\end{algorithmic}
\end{algorithm}
\vspace{-0.8cm}

\subsection{A Low Complexity User Scheduling Algorithm}

The complexity of the user scheduling scheme in \textbf{Algorithm \ref{Iteration}} is strongly related to the channel qualities of terrestrial users, and also scales polynomially with the total number of users $N$ and scales exponentially with the selected number of users $K$ in the worst case. Such a complexity might still not be feasible for power limited terrestrial stations.

In claim (ii) of \textbf{Theorem \ref{bounds}}, we derive a lower bound on the achievable sum rate using $\{S_{u_k}^\mathrm{lb}\}_{k\in [K]}$. Inspired by this idea, we propose a lower-bound based user scheduling algorithm (LBUS) by choosing all users to our theoretical derivations $\{S_{u_k}^\mathrm{lb}\}_{k\in [K]}$ except $u_K^*$. The user scheduling scheme is summarized in \textbf{Algorithm \ref{LB_algorithm}}.

\begin{algorithm}[ht]
\caption{lower-bound based user scheduling algorithm (LBUS)}
\label{LB_algorithm}
\begin{algorithmic}
\Require
$N$, $K$, terrestrial channel coefficients $\{h_{k}\}_{k\in[K]}$, transmission power $P_1$, target rate $R_{\mathrm{target}}$
\Ensure Scheduled users $\bu^*=\{u_1^*,u_2^*,\ldots,u_K^*\}$.
\State Calculate $S_{i}$ for $i\in[K]$, and $S_{\mathrm{max}}$, $S_{\mathrm{min}}$, $\gamma_{\mathrm{target}}$.
\State Choose the first user as $u_1^*=\argmax_{i\in[N]}S_{i}$; Initialize the candidate user set as $\calU=[N]\setminus\{u_1^*\}$.
\State Calculate the candidate set of $K$-th user as $\calU_K:=\{i\in\calU:~S_{K}^{\mathrm{min}}\leq S_{i}\leq S_{K}^{\mathrm{max}}
\}$,
where $S_{K}^{\mathrm{min}}=\hat{S}_{u_K}^{\mathrm{lb}}$, $S_{K}^{\mathrm{max}}=\frac{S_{\mathrm{\mathrm{max}}}}{(\gamma_{\mathrm{target}}+1)^{K-2}\gamma_{\mathrm{target}}}-1$.
\While{$\calU_K\neq\emptyset$}
\State $u_K^*$ is selected by $u_K^*=\argmax_{i\in\calU_K}S_{i}$.
\State Update candidate user set as $\calU$ as $\calU\setminus\{u_K^*\}$
\For{$i=2,\dots, K-1$}
\State User $u_k^*$ is chosen as
\begin{align}
u_k^*=\argmin_{i\in\calU:S_{i}\geq \max\big\{\gamma_{\mathrm{target}}\big(\sum_{i=k+1}^{K}S_{u_i^*}+1\big), S_{u_{k+1}^*} \big\}}S_{i}\label{lselect}.
\end{align}
\State Update the candidate user set as $\calU\setminus\{u_k^*\}$.
\EndFor
\If {\eqref{lselect} does not have a solution for any $k\in[2:K]$}
\State Update the candidate user set as $\calU=[N]\setminus\{u_1^*\}$
\State Update $\calU_K$ as $\calU_K\setminus\{u_K^*\}$.
\Else
\State Break.
\EndIf
\EndWhile
\end{algorithmic}
\end{algorithm}
\vspace{-0.8cm}

\subsection{Numerical Results}
In this section, we present simulation results of our proposed algorithms, i.e., the greedy algorithm GIUS and the lower complexity algorithm LBUS, under the assumption of knowledge of instantaneous CSI for channels. We assume that $\{S_i\}_{i\in[N]}$ are generated i.i.d. from a Rayleigh distribution with the variance $P_1\sigma^2=10$.
To better evaluate the performance of different user scheduling schemes, we assume that the channel between the satellite and the relay is strong enough so that the sum rate is only restricted by user scheduling schemes.
\subsubsection{Achievable Sum Rate}

\begin{figure}[ht]
\centering \mbox{
\subfigure[$N=10$]{
\includegraphics[width=0.5\textwidth,height=0.35\textwidth]{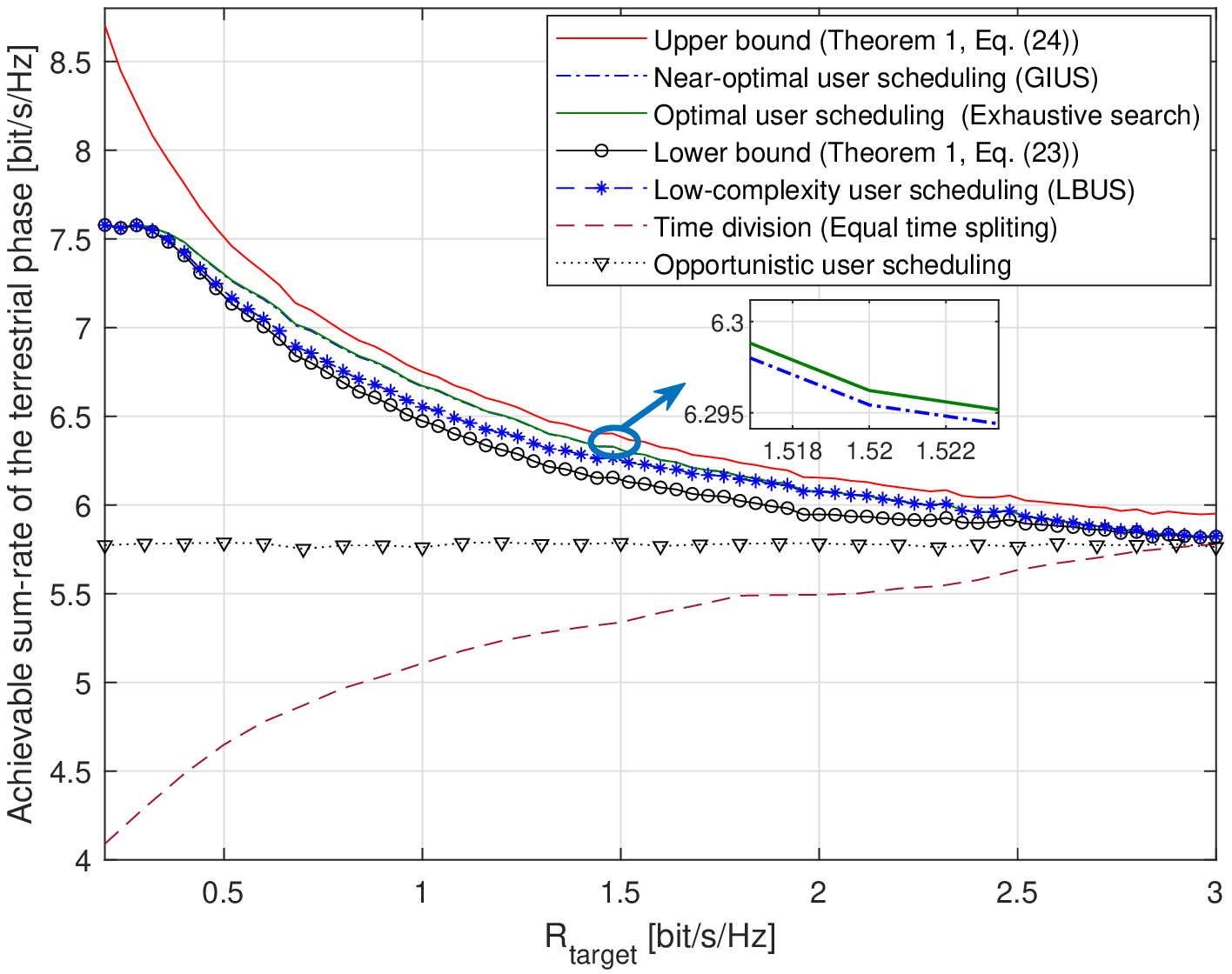}}
\subfigure[$N=20$]{
\includegraphics[width=0.5\textwidth,height=0.35\textwidth]{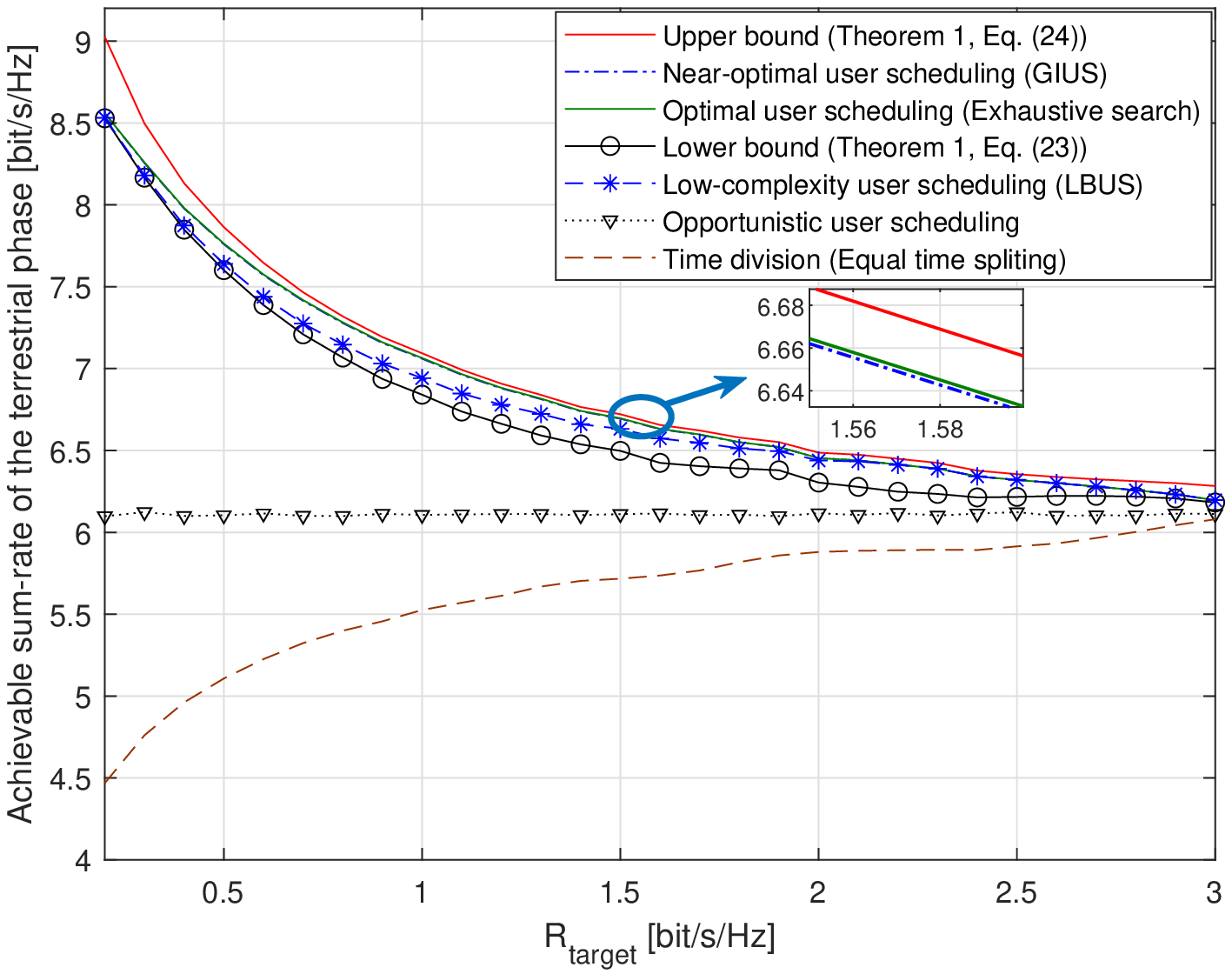}}
}
\caption{Achievable sum rate of user scheduling schemes for different number of users as a function of the target rate $R_{\mathrm{target}}$.}
\label{perfect_CSI}
\end{figure}

In Figure \ref{perfect_CSI}, we plot the simulated performances of our proposed user scheduling algorithms GIUS and LBUS, versus the optimal exhaustive search, a time division multiple access scheme (TDMA) where each ground user access the channel by a equal time duration, an opportunistic user scheduling scheme where the system always selects the user with the strongest channel gain \cite{On_the_Performance}, and our derived upper and lower bounds in \textbf{Theorem \ref{bounds}}. For each user scheduling algorithm, we obtain each data point by averaging the simulated results of $5000$ independent simulations.  Figure \ref{perfect_CSI} shows that our GIUS user scheduling algorithm achieves sum rate with negligible loss compared with exhaustive search and thus demonstrates the efficiency of our GIUS algorithm. Comparing the results in Fig. \ref{perfect_CSI}(a) and (b), we conclude that the gap between the performance of GIUS and the upper bound becomes smaller as the total number of users $N$ increases. This phenomenon implies that GIUS tends to be ``optimal'' for large-scale user groups. It is reasonable because the received SNRs of selected users will tend to the optimal solution $\{S_{u_k}^\mathrm{ub}\}_{k\in[K]}$ when $N$ is large enough.

Besides, Fig. \ref{perfect_CSI} also shows that the performance of LBUS with nearly-linear complexity has worse performance than GIUS, but still better than TDMA and opportunistic user scheduling. This implies that LBUS algorithm, although simple, is still an effective non-orthogonal user scheduling algorithm. Finally, we remark that as $R_{\mathrm{target}}$ increases, the performance of all user scheduling algorithms becomes identical. This is because when $R_{\mathrm{target}}$ is sufficiently large, only one user can be served with the desired target rate and the user scheduling reduces to the orthogonal case.

\subsubsection{Computational Complexity}

In this section, we discuss the computational complexity of our user scheduling algorithms with respect to the total number of users $N$ and the scheduled number of users $K$. From the analysis in Section III-E, we conclude that the user scheduling algorithm LBUS has nearly linear complexity with respect to $N$ as it only needs to select the $K$-th user. The computational complexity of GIUS is more complicated. Without loss of generality, assume that the transmit SNR is normalized as $1$ and assume that the channel from each user to the terrestrial station is subject to Rayleigh fading with the same variance $\sigma^2$, i.e., $h_{i}\sim \mathcal{CN}(0,2\sigma^2\mathbf{I})$ for all $i\in[N]$. For each $k\in[K]$, let the number of candidates of GUIS be $N_k$ and its statistical average with respect to channel statistics is $\mathbb{E}\left( {N_k} \right) = N\int_{S_{\rm{min}}}^{\min \left\{ T(k - 1) - {L_{{\rm{min}}}}(k) - 1,S_{u_{k - 1}^*} \right\}} \frac{1}{2{\sigma ^2}}{e^{ - \frac{x}{2{\sigma ^2}}}}dx$. Specifically, we have
\begin{equation}\label{candidate}
\begin{aligned}
  \mathbb{E}\left( {N_k} \right) &= N\int_{S_{\rm{min}}}^{\min \left\{ T(k - 1) - {L_{{\rm{min}}}}(k) - 1,S_{u_{k - 1}^*} \right\}} \frac{1}{2{\sigma ^2}}{e^{ - \frac{x}{2{\sigma ^2}}}}dx
  \leq N\int_{S_{\rm{min}}}^{  T(k - 1) - {L_{{\rm{min}}}}(k) - 1 } \frac{1}{2{\sigma ^2}}{e^{ - \frac{x}{2{\sigma ^2}}}}\\
  &=N\left( e^{ - \frac{{{S_{{\rm{min}}}}}}{{2{\sigma ^2}}}} - e^{ - \frac{1}{{2{\sigma ^2}}}\left( {T(k - 1) - {L_{\rm{min}}}(k) - 1} \right)} \right)
  \leq N\left( {e^{ - \frac{S_{\rm{min}}}{2{\sigma ^2}}} - e^{ - \frac{1}{{2{\sigma ^2}}}\left( {{S_{\rm{max}}}/{\gamma _{\rm{target}}} - \sum\limits_{i = 2}^{k - 1} {S_{u_{k - 1}^*} - } {L_{\rm{min}}}(k) - 1} \right)}} \right)
  \end{aligned}
\end{equation}
For $N$ large enough, we have $S_{u_{k - 1}^*}\approx S_{u_{k - 1}}^{\mathrm{ub}}$, as the received SNRs of selected users will tend to the theoretical optimal solution given by \eqref{UPB} in this case. If $\frac{S_{\rm{max}}}{\gamma _{\rm{target}}^{k - 2}(\gamma _{\rm{target}}- 1)} - L_{\rm{min}}(k) - 1 > S_{\rm{min}}$, it follows from \eqref{candidate} that
$\mathbb{E}\left( {N_k} \right)\le Ne^{ - \frac{S_{\rm{min}}}{{2{\sigma ^2}}}}\left( 1 - e^{ - \frac{1}{{2{\sigma^2}}}\left({\frac{S_{\rm{max}}}{{\gamma_{\rm{target}}^{k - 2}(\gamma _{\rm{target}} - 1)}} - L_{\rm{min}}(k) - 1 - S_{{\rm{min}}}} \right)} \right)$, which is further upper bounded by
\begin{equation}\label{candidate3}
  \mathbb{E}\left( {N_k} \right) \le Ne^{ - \frac{{{S_{{\rm{min}}}}}}{{2{\sigma ^2}}}}\left( 1 - e^{\frac{1}{2{\sigma ^2}}\left( L_{\rm{min}}(k) + 1 + S_{{\rm{min}}} \right)} \right),
\end{equation}
when $k$ is large enough. Thus, the average number of candidate users $\mathbb{E}\left( {N_k} \right)$ decrease exponentially with $k$, which is the total number of selected users so far. This implies that the set of candidate users shrinks significantly after selecting each user and this is the reason why GIUS has much lower complexity compared to exhaustive search.

\begin{figure}[t]
\centering \mbox{
\subfigure[Computational complexity ]{
\includegraphics[width=0.45\textwidth,height=0.35\textwidth]{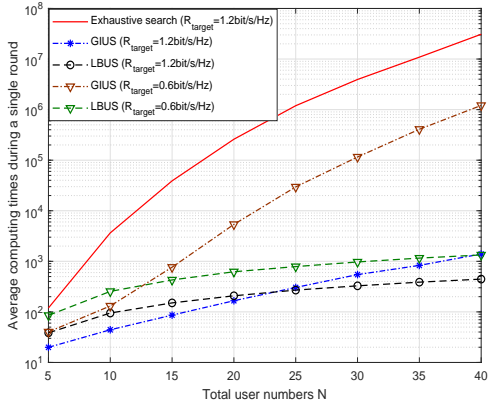}}
\subfigure[Achievable sum rate]{
\includegraphics[width=0.45\textwidth,height=0.35\textwidth]{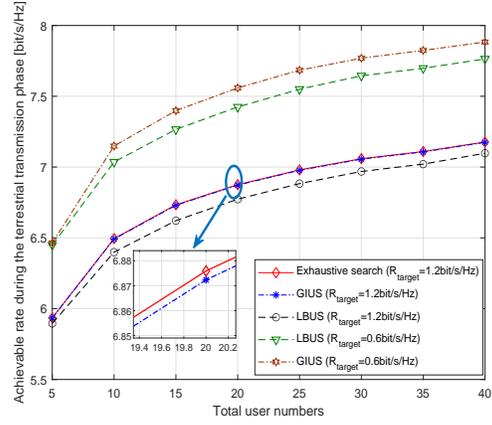}}
} \caption{The complexity and performance comparison between three user scheduling algorithms.} \label{Complexity_running}
\end{figure}

Next, we numerically simulate the complexity of user scheduling algorithms. To do so, we run $2000$ independent simulations of exhaustive search, GIUS, and LBUS algorithms with the target rates $R_{\mathrm{target}}=0.6/1.2 \mathrm{bit/s/Hz}$. We then calculate the average computing times of each user scheduling algorithm. The simulated results are plotted in Figure \ref{Complexity_running}, where Fig. \ref{Complexity_running}(a) demonstrates the running time and Fig. \ref{Complexity_running}(b) plots the achievable sum rate.

Fig. \ref{Complexity_running}(a) shows that when $R_{\mathrm{target}}=1.2 \mathrm{bit/s/Hz}$, the complexity of exhaustive search is much higher than our proposed algorithms GIUS and LBUS. Further, the gaps between the computational complexity of exhaustive search and our algorithms grow when the total number of users grow. On the other hand, Fig. \ref{Complexity_running}(b) shows that for the same case, the achievable sum rates of our GIUS algorithm and exhaustive search are almost the same. This implies that GIUS achieves close-to-optimal performance with much reduced complexity. When the target rate decreases to $R_{\mathrm{target}}=0.6 \mathrm{bit/s/Hz}$, the number of selected users $K$ grow. In this case, it is out of computing power to simulate exhaustive search and we only compare the performance of our proposed algorithms. As expected, LBUS has a much reduced complexity, especially for $K\geq 20$. Surprisingly, the loss of achievable rate is tolerable and implies that our linear complexity algorithm LBUS is very useful when the number of selected users is small.

\begin{figure}[t]
\centering
\begin{minipage}[c]{0.5\textwidth}
\centering
\includegraphics[height=0.6\textwidth]{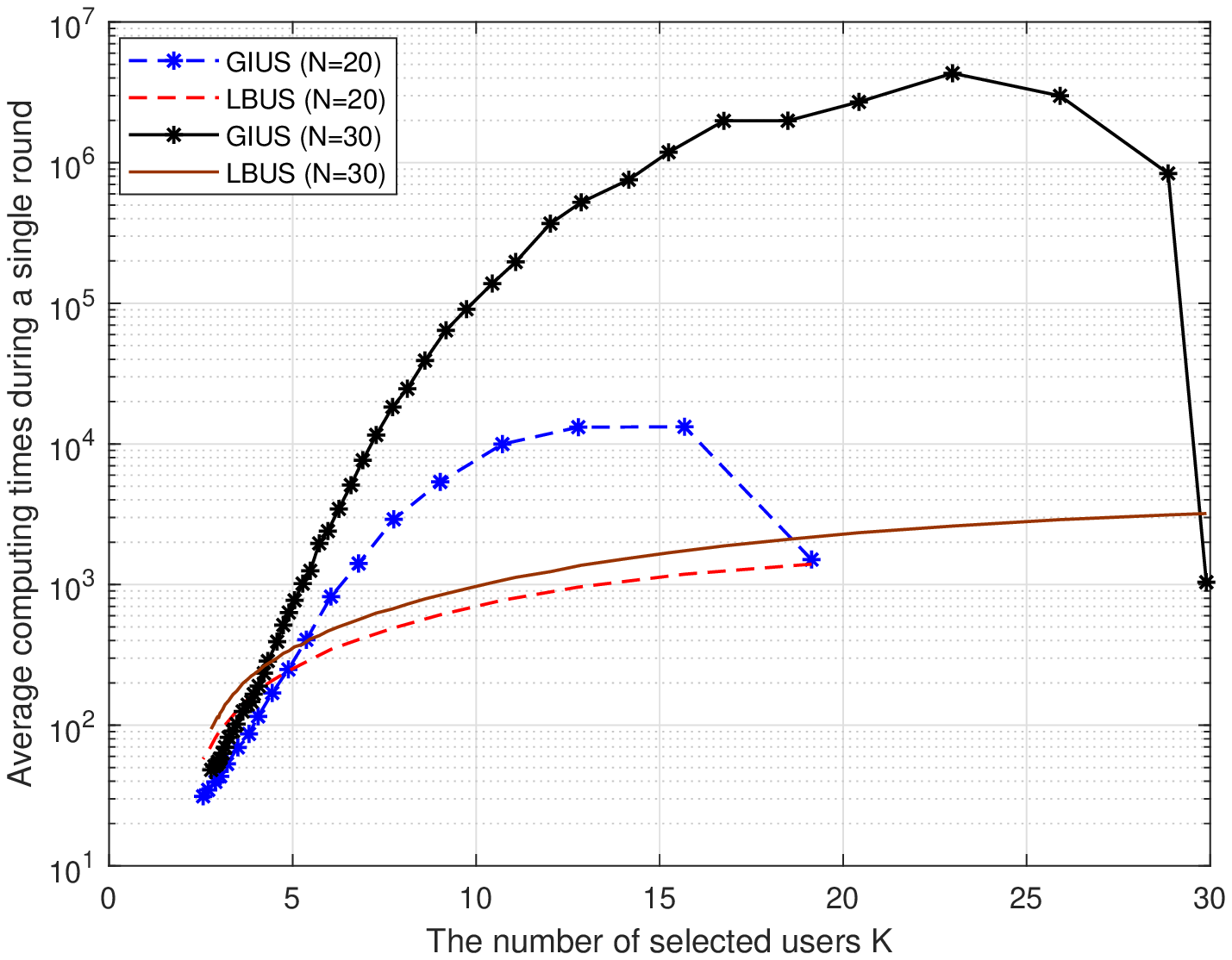}
\caption{The average computing times versus $K$.}\label{Complexity_601}
\end{minipage}%
\begin{minipage}[c]{0.5\textwidth}
\centering
\includegraphics[height=0.6\textwidth]{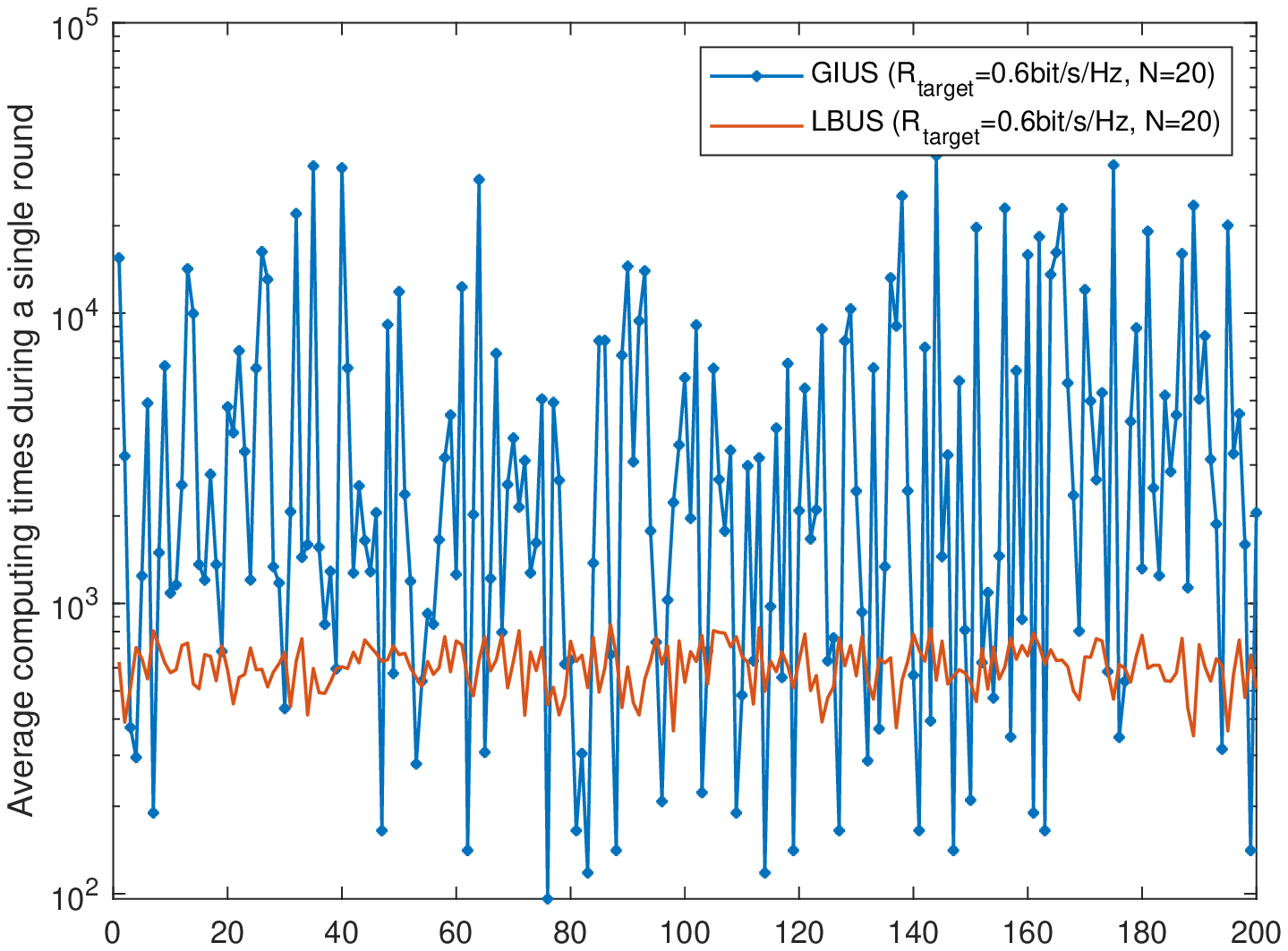}
\caption{The average computing times versus channel variation.}\label{Complexity_variance}
\end{minipage}
\end{figure}

To further compare the computational complexity of our proposed algorithms GIUS and LBUS, we simulate both algorithms for different number of selected users $K$ by changing the target rates $R_{\mathrm{target}}$, when the total number of users $N=40$. The average computing times are plotted in Figure \ref{Complexity_601}. From Figure \ref{Complexity_601}, we find that the computational complexity of GIUS is much more sensitive to the number of selected users $K$ (and thus the target rate $R_{\mathrm{target}}$) while LBUS roughly changes linearly with the number of selected users.
In particular, the average computing times of GIUS grows quickly and reaches its maximum point roughly for $K=20$, while LBUS has much smaller running time for almost all cases.

We also compare the stability of computational complexity of our algorithms. To do so, by setting $N=40$ and $R_{\mathrm{target}}=0.4$ bit/s/Hz, we simulate both algorithms $50$ times independently and record the running time. The results are plotted in \ref{Complexity_variance}. We find that GIUS strongly depends on the channel realizations while LBUS is more stable. This is because the distributions of channels will determine how the searching field shrinks during each iteration of GIUS.

\section{Main Results for the Case with CDI}
\label{Performance_case2}

When the number of users $N$ is large, it is almost impossible to obtain reliable estimates of the fading coefficients for all channels. Thus, we consider the case where only CDI of each channel is available at the relay and the satellite. Under this case, given a target rate, we derive closed-form expressions for the outage probability with two phases of the communication and formulate an optimization problem to minimize the outage probability. We propose a low complexity solution which corresponds to an efficient user scheduling algorithm for terrestrial users.

\subsection{Assumptions and Preliminaries}
The set of all terrestrial users $\calN=[N]$ are divided into $M\leq N$ groups, where users with the same CDI are grouped together. Specifically, for any two users $(i,j)\in[N]_2$, they lie in the same group if the fading coefficients $(h_{i},h_{j})$ for their terrestrial channels to the relay have the same distribution. This is valid since this way, users close in locations are grouped together.

For simplicity, given each $k\in[M]$, we use $\calM_k$ to denote users in group $k$, whose channel fading coefficients follow Rayleigh distribution with parameter $\sigma_k^2$, i.e., $h_{j}\sim \mathcal{CN}(0,2\sigma_k^2\mathbf{I})$ for all $j\in\calM_k$ and $\sigma_k^2$ is the CDI of users in group $\calM_k$. Note that when only CDI is available, user selection is essentially user group selection. For each time slot, $K$ ($K\leq M$) users are selected to transmit messages. Let $\mathbf{w}:=\{w_1,\ldots,w_K\}\in [M]_K$ be the set of the indices of selected user groups and let $\mathbf{v}:=(v_1,\ldots,v_K)\in [N]_K$ be the index vector of the selected users where $v_k\in \mathcal{M}_{w_k}$. Each user $v_k$ in the group $\mathcal{M}_{w_k}$ is selected in a time division manner.

The communication proceeds in the two phases as described in Section II. After user scheduling, $K$ users $\bv$, each from a group are selected. For each $k\in[K]$, user $v_k$ sends a message $M_{v_k}$ to the satellite with power $P_1$ with the aid of the terrestrial relay, who works in full-duplex DF mode with power $P_2$. Successive cancellation decoding is done at the relay and the satellite. The decoding order at the relay is given by $v_1\rightarrow v_2\rightarrow\ldots\rightarrow v_K$, which satisfies $\sigma_{v_1}^2>\sigma_{v_2}^2>\ldots>\sigma_{v_K}^2$. This assumption simplifies the theoretical analysis in the following section. The decoding order at the satellite is related with that at the relay via the bijective mapping $\Psi'\in \mathcal{F}(K,M)$.

\subsection{Expressions of Outage Probabilities}
\label{Problem Formulation outage}
When only CDI is available, we consider the outage probability, which is the probability that the transmission at a given rate is not supported, as the performance criterion. Given $\bv=(v_1,\ldots,v_K)$, the outage event of user $v_j$ occurs if any one of the first $j-1$ user suffers an outage event. Specifically, given a target rate $R_{\mathrm{target}}$ and thus corresponding $\gamma_{\mathrm{target}}$ (cf. \eqref{def:gamma}), for each $k\in[K]$, the outage probability for user $v_k$ during the first phase satisfies
\begin{align}
\mathbb{P}_{v_k,\mathrm{out}}^1 &=1-\mathbb{P}\left( S_{{v_j}}\geq \left(\sum\limits_{i\in[j+1:k]}S_{{v_i}}+1 \right)\gamma _{\mathrm{target}}~,j\in[k]\right),~k\in[K-1],\label{Rout_u_k1}\\
\mathbb{P}_{v_K,\mathrm{out}}^1 &=1-\mathbb{P}\left(
S_{{v_j}}\geq  \left(\sum_{i\in[j+1:k]}S_{{v_i}}+1\right)\gamma _{\mathrm{target}}, j\in [K-1];S_{{v_K}}\geq \gamma _{\mathrm{target}}\right)\label{Rout_u_K},
\end{align}
where $S_{v_k}=P_1|h_{v_k}|^2$ is the received SNR of the signals from user $v_k$.

Recall that DF is used at the relay and a power allocation vector is used to split the power $P_2$ at the relay among $K$ messages of users. Recall that $\gamma_{\Psi(v_k)}$ is the received SNR of user $v_k$ defined in \eqref{rec_relay_SNR2}. For the second phase, the outage probability of each user $v_k$ is
\begin{equation}\label{Rout_u_k_2}
\mathbb{P}_{v_k,\mathrm{out}}^2 =1-\mathbb{P}\left( \gamma_{\Psi(v_k)}\geq\gamma _{\mathrm{target}}~,j\in[k]\right),~k\in[K].
\end{equation}
Note that \eqref{Rout_u_k_2} is subject to the power allocation $\{\alpha_{\psi(v_k)}\}_{k\in[K]}$ and the received SNR $S_{\rmD,\rmR}$ for the satellite-relay channel. From arguments around \eqref{Maximum_K_1}, if the number of selected users satisfies $K\leq \frac{\log(S_{\mathrm{D},\mathrm{R}}+1)}{R_\mathrm{target}}$, we can always find a power allocation scheme that guarantees $\gamma_{\Psi(v_k)}$ is larger than the threshold $\gamma_{\mathrm{target}}$. Thus, the outage probability of each user at the second hop satisfies
$\mathbb{P}_{v_k,\mathrm{out}}^2 =1-\mathbb{P}\left(K\leq \frac{\log(S_{\mathrm{D},\mathrm{R}}+1)}{R_\mathrm{target}}\right)$ for $k\in[K]$, i.e.,
$\mathbb{P}_{v_k,\mathrm{out}}^2=\mathbb{P}_{\mathrm{out}}^2=\mathbb{P}(S_{\mathrm{D},\mathrm{R}}< 2^{KR_{\mathrm{target}}}-1)$ for $k\in[K]$.

In \textbf{Theorem \ref{results:outp}}, we derive a closed-form expression for $\mathbb{P}_{v_K,\mathrm{out}}$, the outage probability for user $v_K$ at the relay. As we shall show, our user scheduling algorithm relies only on $\mathbb{P}_{v_K,\mathrm{out}}$.

\begin{theorem}
\label{results:outp}
The following claims hold.
\begin{enumerate}
\item Given the selected users $\mathbf{v}=\{v_1,v_2,\ldots,v_K\}$ and the corresponding CDI $\{\sigma_{v_k}^2\}_{k\in[K]}$ for their terrestrial channels to the relay, let $\lambda_{v_k}:=\frac{1}{2P_1\sigma_{v_k}^2}$ for $k\in[K]$. The outage probability $\mathbb{P}_{v_K,\mathrm{out}}^1$ of user $v_K$ during the first phase of uplink communication satisfies
\begin{equation}\label{Pout1}
\mathbb{P}_{v_K,\mathrm{out}}^1=1 - \frac{\lambda_{v_K}A_{K - 1}}{\gamma_{\mathrm{target}}B_{K - 1} + \lambda_{v_K}}e^{ - \gamma_{\mathrm{target}}\left((1 + \gamma_{\mathrm{target}})B_{K - 1} + \lambda_{v_K} \right)},
\end{equation}
where
\begin{equation}\label{Pout1_ex}
\begin{aligned}
A_1&=1,B_1=\lambda_{v_1},~
A_k&=\frac{\lambda_{v_k}A_{k-1}}{\gamma_{\mathrm{target}}B_{k-1} + \lambda_{v_k}}, B_k=(1 + \gamma_{target})B_{k - 1} + \lambda_{v_k}, k\in[2,K-1].
\end{aligned}
\end{equation}
\item The outage probability $\mathbb{P}_{\mathrm{out}}^2$ for the second phase communication satisfies
\begin{equation}\label{Pout2}
\begin{split}
\mathbb{P}_{\mathrm{out}}^2
=\left(\frac{2b_0m_s}{2b_0m_s+\Omega}\right)^{m_s}
\cdot\sum_{n=0}^\infty\frac{(m_s)_n}{(n!)^2}\left(\frac{\Omega}{2b_0m_s+\Omega}\right)^n
\Gamma\left(n+1,\frac{2^{KR_{\mathrm{target}}}-1}{2b_0P_2}\right),
\end{split}
\end{equation}
where $(x)_n$ is the Pochhammer symbol and $\Gamma(a,x)=\int_0^xt^{a-1}e^{-t}dt$ is the unnormalized incomplete Gamma function\footnote{The infinite series expression of Eq. \eqref{Pout2} is the series expansion of the confluent hypergeometric function $\sideset{_1}{_1}{\mathop{F}}(a,b,x)$ in \cite{matrix_variates} and \cite{Handbook_of_Mathematical_Functions}. Its convergence and other properties are available in \cite{matrix_variates} and \cite{Handbook_of_Mathematical_Functions}.}.
\item Combining the above two claims, the outage probability for user $u_K$ satisfies $\mathbb{P}_{v_K,\mathrm{out}}=1-(1-\mathbb{P}_{v_K,\mathrm{out}}^1)(1-\mathbb{P}_{\mathrm{out}}^2)$.
\end{enumerate}
\end{theorem}

The proof of \textbf{Theorem \ref{results:outp}} is available in Appendix \ref{proof:outp}.

\subsection{Minimize the Outage Probability}

Note that the outage probability $\mathbb{P}_{\mathrm{out}}^2$ is a fixed function of the SNR of the satellite-relay channel and does not relay on user scheduling in the first phase. Thus, to minimize the outage probability of uplink communication, it suffices to minimize the outage probability for the first phase. We then have the following optimization problem.
\begin{align}
\mathrm{\bf Problem~3:}\quad p_{\mathrm{out}}^* :=&\min_{\substack{ \mathbf{v} \in [N]_K,\mathbf{w} \in [M]_K\\ v_k \in \mathcal{M}_{w_k}, k\in[K]}} \max_{k\in[K]}\{\mathbb{P}_{v_k,\mathrm{out}}\}\label{objective_3}\\
&\mathrm{s.t.}~
\quad \mathbb{P}_{v_k,\mathrm{out}}=1-(1-\mathbb{P}_{v_k,\mathrm{out}}^1)(1-\mathbb{P}_{\mathrm{out}}^2)\label{cp1_3},\\
&\quad \lambda_{v_i}< \lambda_{v_j},~\mathrm{for~}(i,j)\in[K]_2\mathrm{~s.t.~}i<j\label{cp2_3}.
\end{align}
Recall that successive cancellation decoding is used at the relay. Given a decoding order $\bv=(v_1,\ldots,v_K)$, the outage event of user $v_k$ occurs if any user $v_i$ with $i<k$ suffers an outage. Thus, $\mathbb{P}_{v_K,\mathrm{out}}^1\ge \mathbb{P}_{v_k, \mathrm{out}}^1$ for $k\in[K-1]$ and
thus $\max_{k\in[K]}\mathbb{P}_{v_k,\mathrm{out}}=\mathbb{P}_{v_K,\mathrm{out}}$.

\textbf{Problem 3} is an NP-hard problem and it is non-trivial to obtain a closed-form expression for the optimal solution.
Note that the objective function $\mathbb{P}_{v_K,\mathrm{out}}$ of \textbf{Problem 3} depends only on
$\{\lambda_{v_k}\}_{k\in[K]}$. As a compromise, we first derive an lower bound on the objective function by assuming that $\lambda_{v_k}$ is consistent that satisfies $\lambda_{v_k}\geq\lambda_{\mathrm{min}}$, where $\lambda_{\mathrm{min}}:=\min\limits_{k\in[M]}\{\lambda_k\}$ and $\lambda_k:=\frac{1}{2P_1\sigma_k^2}$ for each $k\in[M]$. The relaxed optimization problem is as follows.
\begin{align}
\mathrm{\bf Problem~4:}
&\min\limits_{ \lambda_{v_k}\geq\lambda_{\mathrm{min}}, k\in[K]}\mathbb{P}_{v_K,\mathrm{out}}\\
&\mathrm{s.t.~} \mathbb{P}_{v_K,\mathrm{out}}=1-(1-\mathbb{P}_{v_K,\mathrm{out}}^1)(1-\mathbb{P}_{\mathrm{out}}^2)
\end{align}

\begin{theorem}
\label{results:lambda}
\textbf{Problem~4} has unique optimal solution $\{\lambda_{v_k}^{\rm{op}}\}$, which satisfies the following equalities
\begin{equation}\label{function_series}
  \left\{
  \begin{aligned}
      \lambda_{v_1}^{\rm{op}}&=\frac{1}{2\sigma_{max}^2}, \lambda_{v_K}^{\rm{op}} =-\frac{\gamma _{\mathrm{target}}B_{K-1}}{2} + \frac{\sqrt{\gamma_{\mathrm{target}}^2B_{K-1}^2 + 4B_{K-1}} }{2}\\
\lambda_{v_k}^{\rm{op}}&=1/\left(\sum_{i\in[k:K]}\frac{1}{\lambda_{v_k}^{\rm{op}}+D_{k,i}}+\gamma_{\mathrm{target}}(1+\gamma_{\mathrm{target}})^{K-k}\right),~k\in[2:K-1]
\end{aligned}
\right.
\end{equation}
where $D_{k,i}=\frac{1}{\gamma_{\mathrm{target}}(1+\gamma_{\mathrm{target}})^{i-1-k}}(\gamma_{\mathrm{target}}B_{i-1}+\lambda_{v_i}^{\rm{op}})-\lambda_{v_k}^{\rm{op}}$ for $i\in [k+1:K]$ and $D_{k,k}=\gamma_{\mathrm{target}}B_{k-1}$. Note that $\{D_{k,i}\}_{i\in[k:K]}$ are independent of $\lambda_{v_k}^{\rm{op}}$, and $B_k$ was defined in \eqref{Pout1_ex}.
\end{theorem}
The proof of \textbf{Theorem \ref{results:lambda}} is available in Appendix \ref{proof:lambda}.
\textbf{Theorem \ref{results:lambda}} provides a lower bound to the minimal outage probability via solutions to \textbf{Problem 4} and thus can be used as a benchmark to evaluate the performance of any user scheduling algorithm.

\subsection{A Low Complexity User Scheduling Algorithm}\label{user_scheduling_ave}

\begin{algorithm}[t]
\caption{AO iterative user group selection algorithm (AOIUS)}\label{Iteration_ave}
\begin{algorithmic}
\Require
User group numbers $M$, user group indices $\mathcal{M}_k$ ($k\in [M]$), CDI parameters $\{\sigma_{k}^2\}_{k\in[M]}$, the transmit power $P_1$,
the target SINR $\gamma_{\mathrm{target}}$ , the threshold $\delta$.
\Ensure Scheduled user groups $\mathbf{w}^*=\{w_1^*,w_2^*,\ldots,w_K^*\}$.
\State Calculate $\lambda_{k}=\frac{1}{2P_1\sigma_{k}^2}$ for $k\in[M]$; Choose the first user group as $w_1^*=\arg\min_{\substack{ k\in[M]}}\{\lambda_k\}$.
\State Initialize $\mathcal{W}=\{w_1^*, w_2^0, w_3^0,\ldots, w_K^0\}$ by randomly selecting from $[M]$;
\State Set $\mathbb{P}_{\mathrm{out}}^1(0)=1$. Calculate the outage probability $\mathbb{P}_{\mathrm{out}}^1(1)$ according to (\ref{Pout1}); $i=1$.
\While{$|\mathbb{P}_{out}^1(i)-\mathbb{P}_{\mathrm{out}}^1(i-1)|>\delta$}
\For{$k=2,\dots, K-1$}
\State $\mathcal{W}\gets\mathcal{W}\setminus w_k^{i-1}$.
\State Calculate the zero point $z_{w_k}$ of $h(\lambda_{w_k})$.
\State Calculate $w_k^i$ by $w_k^i =\arg \min_{\substack{j\in[M]\setminus \mathcal{W}\\ \lambda_{w_{k-1}^i}<\lambda_{j}<\lambda_{w_{k+1}^{i-1}}}}|\lambda_{j}-z_{w_k}|$.
\State $\mathcal{W}\gets \mathcal{W}\cup\{w_k^i\}$.
\State Update the value of $\{B_j\}_{j\in[k:K]}$ according to the current $\mathcal{W}$ based on (\ref{Pout1}).
\EndFor
\State $\mathcal{W}\gets\mathcal{W}\setminus w_K^{i-1}$.
\State Calculate $w_K^i$ by $w_K^k=\arg\min_{\substack{j\in[M]\setminus \mathcal{W}\\ \lambda_{w_{K-1}^i}<\lambda_{j}<\lambda_{\mathrm{max}}}}|\lambda_{j}-z_{w_K}|$, where $z_{w_K}=-\frac{\gamma_{\mathrm{target}}B_{K - 1}}{2} + \frac{\sqrt{\gamma_{\mathrm{target}}^2B_{K - 1}^2 + 4B_{K - 1}} }{2}$.
\State $\mathcal{W}\gets \mathcal{W}\cup\{w_K^i\}$.
\State $i\gets i+1$. Calculate $\mathbb{P}_{\mathrm{out}}^1(i)$ and $B_j$ ($j\in[K]$) based on \eqref{Pout1}.
\EndWhile
\State $\mathbf{w}^*\gets\mathcal{W}$.
\end{algorithmic}
\end{algorithm}

Note that theoretical solutions in \eqref{function_series} are higher degree equations and don't have closed-form expressions.
In this section, to minimize the outage probability, we propose a solution to \textbf{Problem 3} which corresponds to a user scheduling algorithm.
Recall that for each $k\in[M]$, $\sigma_k^2$ is the CDI for users in group $\calM_k$ and $\lambda_k=\frac{1}{2P_1\sigma_k^2}$.
We define a function $h(\lambda_{w_k})=\frac{1}{\lambda_{w_k}}-\sum_{i\in[k:K]}\frac{1}{\lambda_{w_k}+D_{k,i}}-\gamma_{\mathrm{target}}(1+\gamma_{\mathrm{target}})^{K-k}$ for each $k\in[2,K-1]$.

Note that \textbf{Theorem 3} provides an optimal solution $\{\lambda_{v_k}^{\rm{op}}\}_{k\in[K]}$ to \text{\bf Problem 4}, a relaxed version of \textbf{Problem 3}. To obtain a feasible solution, our user scheduling algorithm proceeds by choosing the $\lambda_{v_k}$ close to $\lambda_{v_k}^{\rm{op}}$.
As stated before, the user selection is essentially user group selection.

We select user groups $\bw^*=(w_1^*,\ldots,w_K^*)$ as follows. The first user group $w_1^*$ is selected as the one with the maximal CDI $\sigma_k^2$, i.e., $w_1^*=\arg\min_{\substack{ k\in[M]}}\{\lambda_k\}$. When only two users are allowed, the second one is selected as $w_2^*=\arg\min_{\substack{k\in[M]\setminus w_1^*}}|\lambda_k-z_{w_2}|$, where
$z_{w_2}:=-\frac{\gamma _{\mathrm{target}}\lambda_{w_1^*}}{2} + \frac{\sqrt{\gamma_{\mathrm{target}}^2\lambda_{w_1^*}2 + 4\lambda_{w_1^*}} }{2}$.

When more than two users are allowed, i.e., $K>2$, the selection of user groups is more involved and we propose an iterative user group selection scheme based on alternate optimization that is summarized in \textbf{Algorithm \ref{Iteration_ave}} and named AOIUS.
We remark that AOIUS converges because the outage probability monotonically decreases during iteration rounds. Specifically, for $i\in\bbN$, in the $i$-th iteration, for each $k\in[2:K]$, the $k$-th user (group) is selected to minimize the outage probability, which is calculated with respect to parameters $\left\{\sigma_{w_j^i}^2\right\}_{j\in[1:k-1]}$ updated in the current iteration and $\left\{\sigma_{w_j^{i-1}}^2\right\}_{j\in[k+1:K]}$ obtained in the last iteration. Therefore, the outage probability converges quickly after the first round of the iteration, and slows down as the iteration progresses.

\subsection{Numerical Results}
To demonstrate the efficiency of our proposed algorithm, we systematically simulate the performance of AOIUS in terms of outage probability and computational complexity. The channel between the relay and the satellite are assumed to be under frequent heavy shadowing, whose parameters are given by $(b_0,m_s,\Omega)=(0.063,0.739,8.97\times10^{-4})$ \cite{Land_Mobile_Satellite}.
In order to eliminate random effects, we assume that $P_1\sigma_k^2$ of group $k$ is uniformly distributed between $(-\infty, 20\text{ dB}]$, based on which average outage probability is computed after 200 times of average channel parameters realizations.
The transmit power at the relay is given by $P_2=30\text{ dB}$.
Note that the effects of the noise and large-scale fading of the terrestrial and the satellite channels are not considered during simulation, as they can be included in the transmit power $P_1$ and $P_2$ and normalized.

\subsubsection{Convergence}

To illustrate the convergence of our user scheduling algorithm AOIUS, we simulate AOIUS and plot the results in Fig. \ref{convergence}.  Without loss of generality, assume that the channel gain between the satellite and the relay is strong enough so that $\mathbb{P}_{out}^2\approx 1$.
This assumption doesn't affect the convergence analysis.
The terrestrial average received SNR $P_1\sigma_{k}^2$ of group $k$ is assumed uniformly distributed over $(-\infty, 20 \text{ dB}]$. The target rate is set as $R_{\mathrm{target}}=0.02$bit/s/Hz and the total number of user groups $M$ is assumed to be $7000$. The threshold $\delta$ for the stopping criterion is set as zero, which means the algorithm stops only when the final outage probability doesn't change. We simulate AOIUS when the number of allowed users $K=5,10,15$. For each $K$, we simulate AOIUS five times for independent channel realizations and plot calculated outage probabilities versus iterations. From Fig. \ref{convergence}, we find that using AOIUS, the outage probability of scheduled users monotonically decreases with iteration times and finally converges to a fixed value, which is the theoretical lower bound on the outage probability obtained from \textbf{Theorem \ref{results:lambda}}. This implies that our user scheduling algorithm is close to optimal in this numerical example.

\subsubsection{Outage Probability}

\begin{figure}[t]
\centering
\begin{minipage}[c]{0.5\textwidth}
\centering
\includegraphics[height=0.7\textwidth]{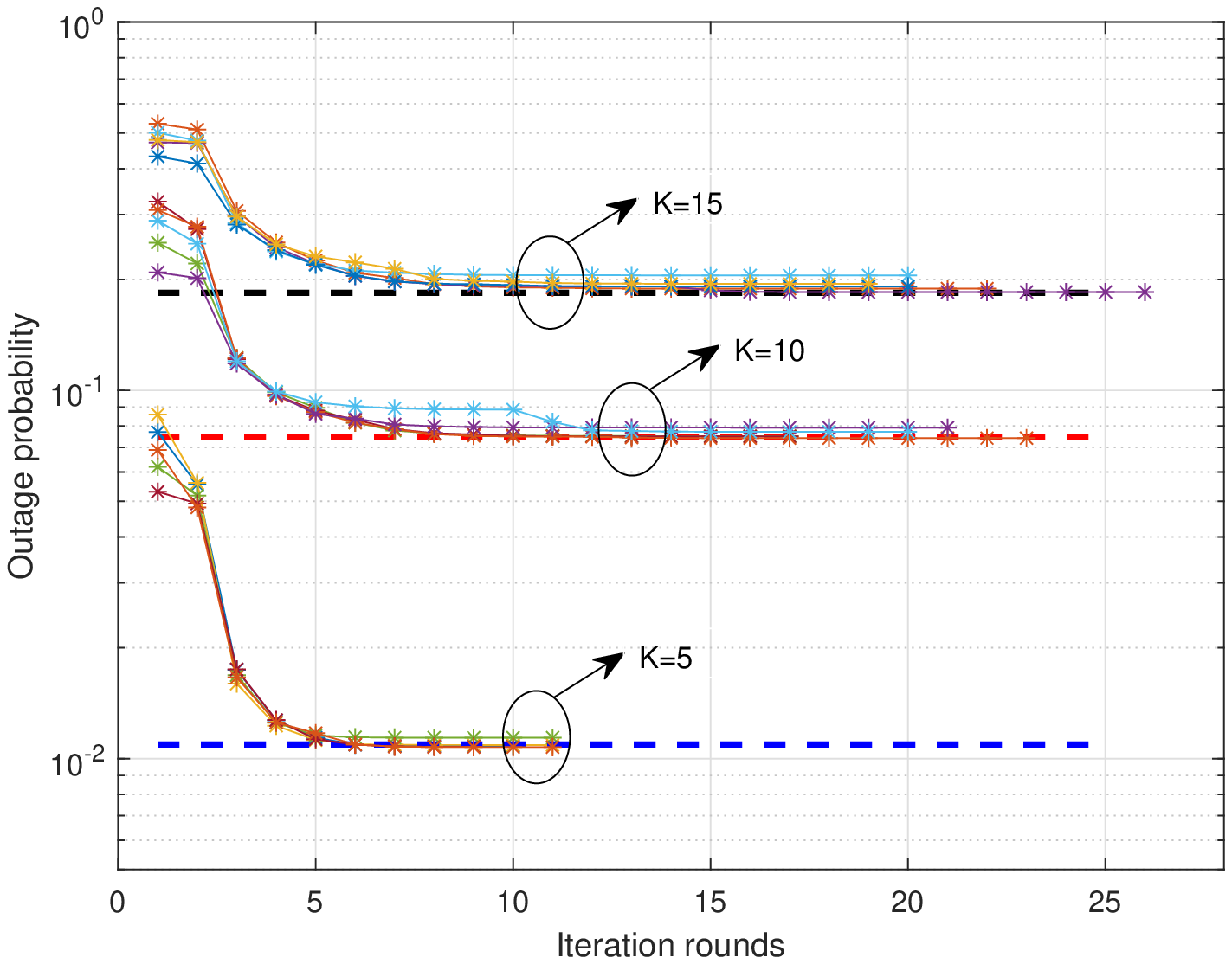}
\caption{Illustration of the quick convergence of AOIUS.}\label{convergence}
\end{minipage}%
\begin{minipage}[c]{0.5\textwidth}
\centering
\includegraphics[height=0.7\textwidth]{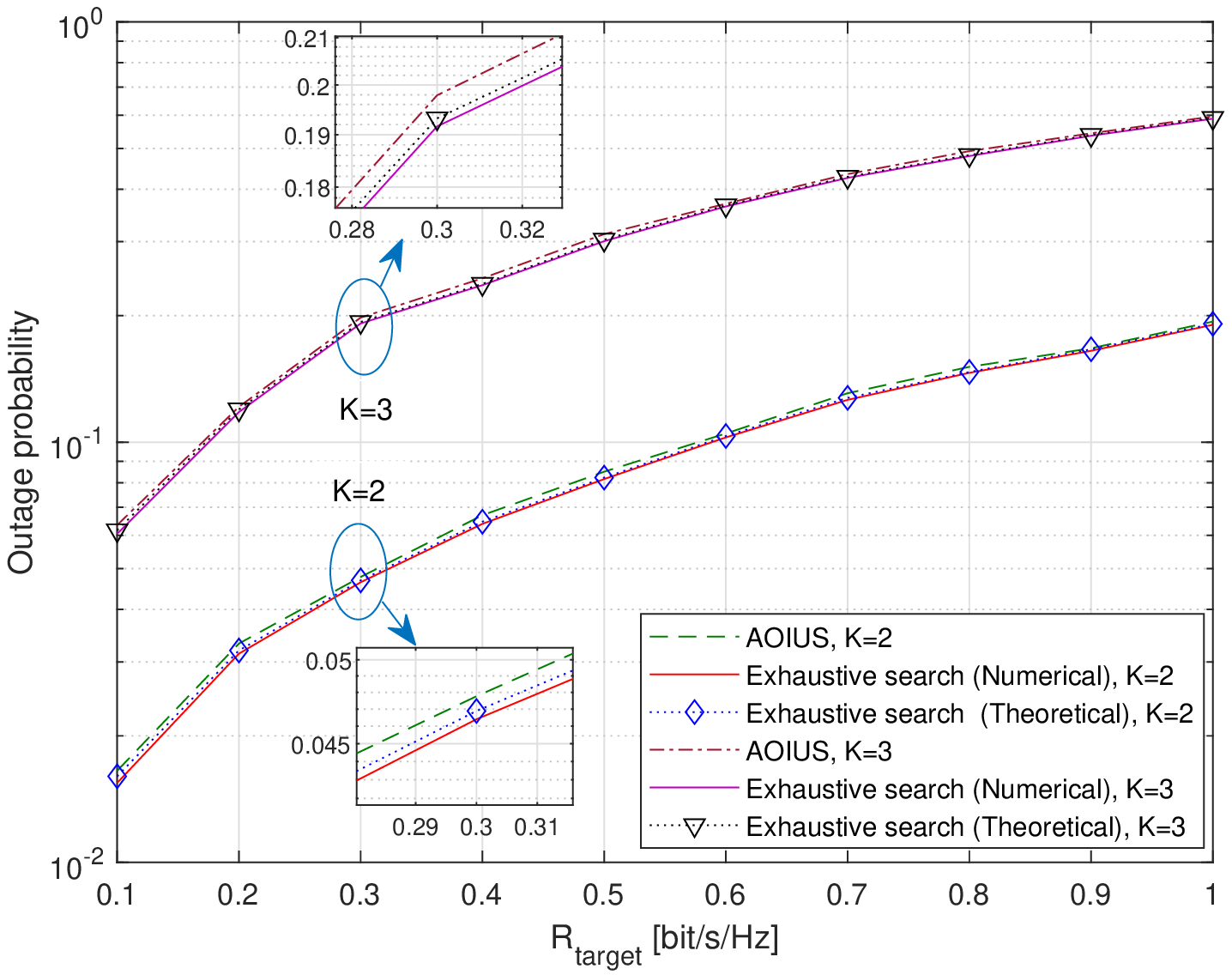}
\caption{Simulated Outage probabilities.}\label{Outage_total}
\end{minipage}%
\end{figure}

To demonstrate the efficiency of our proposed algorithm, we simulate AOIUS for the cases of $K=2, 3$ and compare the results with the optimal exhaustive search strategy. We set $M=10$, and $\delta=0$ as the stopping criterion. The results are plotted in Fig. \ref{Outage_total}. The difference between theoretical and simulation results is that the theoretical results correspond to simulations of user schedule algorithms where the outage probability is calculated from \eqref{Pout_K_ana} and numerical correspond to Monte Carlo simulation of the outage probability, which is calculated as the ratio of outage event for $10^4$ independent realizations of fading channels. For both user scheduling algorithms, the matching of theoretical and numerical simulation results verify the correctness of the calculation of the outage probability in \eqref{Pout_K_ana}. Furthermore, since exhaustive search is optimal, the almost negligible gap between the outage probabilities of AOIUS and exhaustive search implies that our proposed user scheduling algorithm is efficient in minimizing the outage probability. Finally, we comment that the outage probability increases with number of selected users $K$. This is because interference among users is more severe when more users transmit messages simultaneously.

\subsubsection{Computational Complexity}

\begin{figure}[t]
\centering
\begin{minipage}[c]{0.5\textwidth}
\centering
\includegraphics[height=0.6\textwidth]{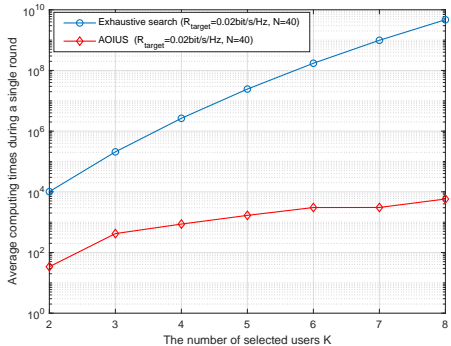}
\caption{Computing times of exhaustive search and AOIUS.}\label{RunningTime}
\end{minipage}%
\begin{minipage}[c]{0.5\textwidth}
\centering
\includegraphics[height=0.6\textwidth]{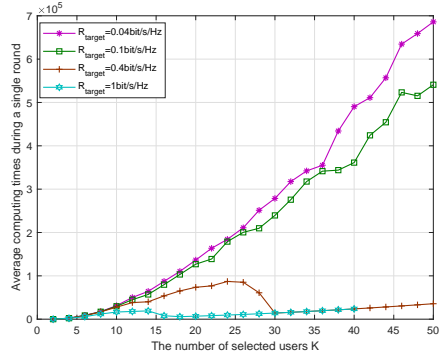}
\caption{Computing times of AOIUS.}\label{RunningTime_iter}
\end{minipage}
\end{figure}

The optimal exhaustive search is not practical since its complexity scales exponentially as the number of selected users grows. The complexity of AOIUS depends on channel distribution, the target rate, the total number of user groups $M$ and the number of selected user groups $K$. To illustrate the complexity of AOIUS, we simulate the computing times of AOIUS for different $K$ and target rates. The complexities of AOIUS and exhaustive search are shown in Fig. \ref{RunningTime}, where each data point corresponds to the average computing time of a single run of user scheduling algorithms over 100 realizations of $\{\sigma_{i}\}_{i\in[M]}$. In this case, we set $M=40$ and $R_{\mathrm{target}}=0.02$bit/s/Hz. From Fig. \ref{RunningTime}, we find that the complexity of AOIUS is much lower and stable, especially for a large number of selected users $K$.

Finally, to explore the effect of the target rate $R_{\mathrm{target}}$ and the selected number of user groups $K$ on the computational complexity of AOIUS, we simulate AOIUS for $M=5000$ user groups and target rates $R_{\mathrm{target}}=0.04/0.1/0.4/1$bit/s/Hz respectively and plot the results in Fig. \ref{RunningTime_iter}. From Fig. \ref{RunningTime_iter}, we observe that the computation complexity of AOIUS roughly grows linearly with the number of selected users when $R_{\mathrm{target}}=0.04/0.1$bit/s/Hz. However, when $R_{\mathrm{target}}=0.4/1$bit/s/Hz, the computational complexity does not scale much with $K$. This is because when the target rate is relatively high, the outage event occurs frequently.
In extreme case, AOIUS stops at the very beginning of the algorithm as the outage probability is always equals to 1. Thus, the computational complexities for the cases of $R_{\mathrm{target}}=0.4/1$bit/s/Hz  drop down when $K$ is larger than a certain threshold (denoted by $K_0$), and then grows very slowly as $K$ tends to infinity. This implies that the system cannot accommodate as more than $K_0$ users with the desired target rate. However, the exact characterization of $K_0$ is challenging and left as future work.

\section{Conclusion}
We studied the hybrid uplink satellite-terrestrial communication model with a single satellite, a terrestrial station as a relay and finitely many ground users. Under mild conditions on the channel models, we studied user scheduling algorithms to maximize the sum rate when perfect CSI of all channels are available and to minimize the outage probability when only CDI of all channels are available. Specifically, when perfect CSI is available, we first derived theoretical upper and lower bounds on the instantaneous sum rate and then proposed a greedy iterative user scheduling algorithm that achieves a near optimal performance with lower complexity. Furthermore, we also proposed a user scheduling algorithm based on our lower bound that has worse performance than the greedy algorithm but also achieves much lower complexity, especially if one wishes to schedule a large number of users in a time slot. When only CDI is available, we first derived analytical expressions for the outage probabilities of two-phase communication and then proposed an alternative optimization based user scheduling algorithm that strikes a balance between performance and complexity. Our results in this paper could shed lights on and inspire the design of user scheduling algorithms for uplink communication of hybrid satellite-terrestrial models.

There are several avenues for future research. Firstly, one can generalize our results to the case where the satellite, the relay and even all the grounds users are equipped with multiple antennas. To do so, one need to use corresponding information theoretical results for the multiple input and multiple out channels~\cite{Wang2005relay,goldsmith2003jsac}. Secondly, one could generalize our results to the more practical communication models with multiple terrestrial relays and multiple satellites. Such analysis requires novel ideas of terrestrial station selection from the perspective of the satellite. Finally, one can also consider block fading models and derive the corresponding results for both CSI and CDI cases~\cite{yang2013jsac,austin2019tit}.

\appendix
\subsection{Proof of Theorem \ref{bounds}}
\label{pf:bounds}
Note that the optimizing over $\mathbf{u}$ is equal to that over SNRs since we select users and decoding orders using SNR.

\begin{enumerate}
\item The if part is easy. If \eqref{claim1} holds, $\{\hat{S}_{u_k}^{\mathrm{lb}}\}_{k\in[K]}$ is a feasible solution and \textbf{Problem 2(a)} is feasible. Conversely, the only if part is proved through the following two steps. Firstly, we show that if \textbf{Problem 2(a)} is feasible, then any feasible solution $\{S_{u_k}\}_{k\in[K]}$ satisfies $S_{u_k}\geq \hat{S}_{u_k}^{\mathrm{lb}}$ for each $k\in[K]$, which implies that $\{\hat{S}_{u_k}^{\mathrm{lb}}\}_{k\in[K]}$ lower bounds any feasible solutions to \textbf{Problem 2(a)}.  Secondly, we use the above lower bound claim to conclude a contradiction if \eqref{claim1} fails to hold and \textbf{Problem 2(a)} is feasible.

    We first show that $\{\hat{S}_{u_k}^{\mathrm{lb}}\}_{k\in[K]}$ lower bounds any feasible solutions.
If $\{S_{u_k}\}_{k\in[K]}$ is feasible to \textbf{Problem 2(a)} but satisfies $S_{u_i}<\hat{S}_{u_i}^{\mathrm{lb}}$ for some $i\in [K]$, we have
\begin{align}\label{impossible}
\gamma_\mathrm{target}&\leq S_{u_K}<\phi(\gamma_\mathrm{target},\mathcal{S})\text{~if~}i=K;\\
\gamma_{\mathrm{target}}\left(\sum_{j\in[i+1:K]}\hat{S}_{u_j}^{\mathrm{lb}}+1\right)
&\leq S_{u_i}<
\phi\left(\gamma_{\mathrm{target}}\left(\sum_{j\in[i+1:K]}\hat{S}_{u_j}^{\mathrm{lb}}+1\right),
\mathcal{S}\setminus\{\hat{S}_{u_t}^{\mathrm{lb}}\}_{t\in[i+1,K]}\right)\text{~if~}i\in[K-1]
\end{align}
However, from the definition of $\phi(\cdot)$, no such $S_{u_i}$ exists. Therefore, any feasible solution $\{S_{u_k}\}_{k\in[K]}$ satisfies $S_{u_k}\geq \hat{S}_{u_k}^{\mathrm{lb}}$ for $k\in[K]$.

Secondly, if \eqref{claim1} fails to hold, then either
$\gamma_{\mathrm{target}}>S_{\mathrm{max}}$ or there exists an index $i\in[K-1]$ that satisfies $\gamma_{\mathrm{target}}\left(\sum_{j\in[i+1:K]}\hat{S}_{u_j}^{\mathrm{lb}}+1\right)> S_{\mathrm{max}}$. For the former case, if \textbf{Problem 2(a)} is feasible, then from our claim above, each feasible solution $\{S_{u_k}\}_{k\in[K]}$ satisfies that $S_{u_K}\geq S_{u_K}^{\mathrm{lb}}>S_{\mathrm{max}}$. This is impossible from the definition of $S_{\mathrm{max}}$. For the latter case, if \textbf{Problem 2(a)} is feasible, then our above claim implies that $\hat{S}_{u_i}^{\mathrm{lb}}>\gamma_{\mathrm{target}}\left(\sum_{j\in[i+1:K]}\hat{S}_{u_j}^{\mathrm{lb}}+1\right)$ and each feasible solution $\{S_{u_k}\}_{k\in[K]}$ satisfies $S_{u_i}\geq S_{u_i}^{\mathrm{lb}}>S_{\mathrm{max}}$. Again, this is impossible.

Therefore, we have now showed that \textbf{Problem 2(a)} cannot be feasible if \eqref{claim1} fails to hold and thus the only if part is proved.

\item The lower bound follows since the chosen user scheduling vector $\bu$ corresponding to $\{S_{u_k}^{\mathrm{lb}}\}_{k\in[K]}$  is a feasible solution to \textbf{Problem 2(a)}.

\item The upper bound is obtained by relaxing the constraints in \textbf{Problem 2(a)} and solving the following new linear programming problem
\begin{align}
\mathrm{\bf Problem~2(a)~LP:~}
&\max_{\substack{S_{u_k}\leq S_{\mathrm{max}}}} C\left(\sum_{k\in[K]}S_{u_k}\right)\\
&\mathrm{s.t.~} C(\gamma_{u_k})\geq R_{\rm{target}}, k\in[K],\\
&~\quad S_{u_i}> S_{u_j},~(i,j)\in[N]_2\mathrm{~and~}i\neq j,
\end{align}
where the optimization parameters $S_{\mathrm{R},u_k}$ are assumed continuous and satisfy $S_{u_k}\leq S_{\mathrm{max}}$. Note that claim (iii) holds since the optimization region of SINRs is enlarged.
\end{enumerate}

\subsection{Proof of Theorem \ref{results:outp}}
\label{proof:outp}
We only provide proofs for claim (i) and (ii) in \textbf{Theorem 2}, as the last claim is straightforward.
\begin{enumerate}
  \item For Rayleigh fading channels, the probability dense function (pdf) of the received SNR $S_{R,k}$ is $P_{S_{k}}(x)=\lambda_ke^{-\lambda_k x}$, where $\lambda_k=\frac{1}{2P_1\sigma_k^2}$.
      Using the properties of exponential distribution, we can derive the close-form of the outage probability during the first hop as
\begin{equation}\label{Pout_K_ana}
\begin{split}
  \mathbb{P}_{u_K,\mathrm{out}}^1&=1-\oiint\limits_{\substack{
x_K \ge \gamma _{\mathrm{target}}\\
x_{K - 1} \ge \gamma _{\mathrm{target}}(x_K + 1)\\
\ldots\\
x_1 \ge \gamma _{\mathrm{target}}(x_2 + x_3 + ,..., + x_K + 1)}}
 p_{S_{R,u_{K}},S_{R,u_{K - 1}},\ldots,S_{R,{u_1}}}(x_K,x_{K - 1},...,x_1)d{x_K}d{x_{K - 1}},....,d{x_1}\\
 &=1-\oiint\limits_{\substack{
x_K \ge \gamma _{\mathrm{target}}\\
x_{K - 1} \ge \gamma _{\mathrm{target}}(x_K + 1)\\
\ldots\\
x_1 \ge \gamma _{\mathrm{target}}(x_2 + x_3 + ,..., + x_K + 1)}}
 \left( \prod\limits_{j = 1}^K \lambda _{u_j}  \right)e^{ - \sum\limits_{i = 1}^K \lambda _{u_i}x_{u_i}}d{x_K}d{x_{K - 1}},....,d{x_1}\\
 &=1 - \frac{\lambda_{u_K}A_{K - 1}}{\gamma_{\mathrm{target}}B_{K - 1} + \lambda_{u_K}}e^{ - \gamma_{\mathrm{target}}\left((1 + \gamma_{\mathrm{target}})B_{K - 1} + \lambda_{u_K} \right)}
\end{split}
\end{equation}

  \item The outage probability during the second transmission phase is given by $\mathbb{P}_{\mathrm{out}}^2=\mathbb{P}(S_{\mathrm{D},\mathrm{R}}<2^{KR_{\mathrm{target}}}-1)
      =\int_0^{2^{KR_{\mathrm{target}}}-1}p_{S_{\mathrm{D},\mathrm{R}}}(s)ds$,
      where $p_{S_{\mathrm{D},\mathrm{R}}}(s)$ is the pdf of $S_{\mathrm{D},\mathrm{R}}$.
      Using \cite{Land_Mobile_Satellite} (Eq. (3)), the pdf of $S_{\mathrm{D},\mathrm{R}}$ is derived as \eqref{pdf_SR}
\begin{equation}\label{pdf_SR}
  p_{S_{\mathrm{D},\mathrm{R}}}(s)=\frac{1}{2P_2b_0}\left(\frac{2b_0m_s}{2b_0m_s+\Omega}\right)^{m_s}
  \exp\left(-\frac{s}{2P_2b_0}\right)\sideset{_1}{_1}{\mathop{F}}\left(m_s,1,\frac{\Omega s}{2P_2b_0(2b_0m_s+\Omega)}\right)
\end{equation}
       where $\sideset{_1}{_1}{\mathop{F}}(a,b,x)$ is the confluent hyper-geometric function \cite{Handbook_of_Mathematical_Functions}. The proof is done by using the fact that $\sideset{_1}{_1}{\mathop{F}}(a,b,x)$ can be simplified as $\sideset{_1}{_1}{\mathop{F}}(a,b,x)=\sum_{n=0}^\infty\frac{(a)_n}{n!(b)_n}x^n$~\cite{Shadowed-Rice_Random_Variables}.
\end{enumerate}

\subsection{Proof of Theorem \ref{results:lambda}}
\label{proof:lambda}

Note that $\mathbb{P}_{u_K,\mathrm{out}}$ is a function of $\lambda_{u_1},\ldots,\lambda_{u_K}$. For subsequent analysis, let $f(\lambda_{u_1},\ldots,\lambda_{u_K}):=\log(1-\mathbb{P}_{u_K,\mathrm{out}}^1)$. This way, the minimization of  $\mathbb{P}_{u_K,\mathrm{out}}$  is equivalent to the maximization of $f(\cdot)$.

The partial derivation of $f(\cdot)$ with respect to $\lambda_{u_k}$ satisfies
\begin{equation}\label{df}
\begin{split}
  \frac{\partial f(\cdot)}{\partial \lambda_{u_k}}&=
  \frac{\gamma_{\mathrm{target}}B_{k - 1}}{\lambda_{u_k}(\gamma_{\mathrm{target}}B_{k - 1} + \lambda_{u_k})}
   - \left(\gamma_{\mathrm{target}}(\gamma_{\mathrm{target}} + 1)^{K - k} + \sum\limits_{i = k + 1}^K \frac{\gamma_{\mathrm{target}}(\gamma_{\mathrm{target}} + 1)^{i - k - 1}}{\gamma_{\mathrm{target}}B_{i - 1} + \lambda_{u_i}}  \right)\\
  &=\frac{1}{\lambda_{u_k}}-\sum_{i\in[k:K]}\frac{1}{\lambda_{u_k}+D_{k,i}}
  -\gamma_{\mathrm{target}}(1+\gamma_{\mathrm{target}})^{K-k}, k\in[2,K-1].
\end{split}
\end{equation}
We then prove that \eqref{df} only has one zero point, which implies $f(\cdot)$ only has one global maximum value. The second partial derivative of $f(\cdot)$ with respect to $\lambda_{u_k}$ for each $k\in[2,K-1]$ is
$\frac{\partial^2 f(\cdot)}{\partial \lambda_{u_k}^2}= -\frac{1}{\lambda_{u_k}^2}+\sum_{i\in[k:K]}\left(\frac{1}{\lambda_{u_k}+D_{k,i}}\right)^2=\frac{g(\lambda_{u_k})}{\lambda_{u_k}^2}$,
where $g(\lambda_{u_k}):=-1+\sum_{i\in[k:K]}\left(\frac{\lambda_{u_k}}{\lambda_{u_k}+D_{k,i}}\right)^2$.
Note that $g(\lambda_{u_k})$ monotonically increases with $\lambda_{u_k}$, and the extreme values satisfy $\lim_{\lambda_{u_k}\rightarrow 0}\sum_{i\in[k:K]}g(\lambda_{u_k})=-1$, $\lim_{\lambda_{u_k}\rightarrow +\infty}g(\lambda_{u_k})=K-k$. Thus, there only exists one zero point $z_0$ for $\frac{\partial^2 f(\cdot)}{\partial \lambda_{u_k}^2}$ such that $\frac{\partial^2 f(\cdot)}{\partial \lambda_{u_k}^2}\geq0$ if and only if $\lambda_{u_k}\geq z_0$, i.e., $\frac{\partial f(\cdot)}{\partial \lambda_{u_k}}$ first decreases in $\lambda_{u_k}$ and then increases in $\lambda_{u_k}$ with the critical point of $z_0$. Further, we have $\lim_{\lambda_{u_k}\rightarrow 0}\frac{\partial f(\cdot)}{\partial \lambda_{u_k}}=+\infty$ and $\lim_{\lambda_{u_k}\rightarrow +\infty}\frac{\partial f(\cdot)}{\partial \lambda_{u_k}}=-\gamma_{\mathrm{target}}(1+\gamma_{\mathrm{target}})^{K-k}$. Thus, there is only one zero point $\lambda_{u_k}^{\rm{op}}$ of the first derivative $\frac{\partial f(\cdot)}{\partial \lambda_{u_k}}$ and $f(\cdot)$ increases in its parameters first and then decreases with the critical point $\lambda_{u_k}^{\rm{op}}$. As a result, $f(\cdot)$ has only one global maximum value for any $k\in[2,K-1]$. The cases of $k=1,K$ are straightforward so the detailed analysis are omitted.
The solution to \textbf{Problem 4}, $\lambda_{u_k}^{\rm{op}}$ ($k\in [K]$) meets \eqref{function_series}. The theorem is proved.

\bibliographystyle{IEEEtran}
\bibliography{BiB}

\end{document}